\documentclass[iop,numberedappendix]{emulateapj}

\usepackage{epstopdf}
\usepackage{graphics}
\usepackage{natbib}

\newcommand{\greeksym}[1]{{\usefont{U}{psy}{m}{n}#1}}

\newcommand{\mum}{\mbox{\greeksym{m}m}}

\newcommand{\Msol}{\mbox{$M_{\odot}$}}
\newcommand{\msun}{\mbox{$\rm M_{\odot}$}~}

\newcommand{\lsun}{\mbox{$L_{\odot}$}~}
\newcommand{\kms}{\mbox{km s$^{-1}$}}

\newcommand{\etal}[1]{{ et al.}~}
\def\kms{\ifmmode \hbox{km~s}^{-1}\else km~s$^{-1}$\fi}
\def\etal {{\it et al.}}
\def\deg      {{\ifmmode^\circ\else$^\circ$\fi} } 

\def\h2     {H$_2$}

\def\h2     {H$_2$}

\def\arcsec{\hbox{$^{\prime\prime}$}}

\begin{document}

\shorttitle{Evolution of Gas, and Star Formation from z = 0 to 5}
\shortauthors{Scoville \etal}

\title{Cosmic Evolution of Gas and Star Formation} \footnote[0]{2022 AAS Russell Lecture by Scoville.
We provide the science background behind the Russell Lecture and draw extensively on previous work in Scoville \etal (2017).
\medskip}

\medskip

 \author{ Nick Scoville\altaffilmark{1}\\ 
 Andreas Faisst \altaffilmark{2}, John Weaver \altaffilmark{3}, Sune Toft\altaffilmark{3}, Henry J. McCracken\altaffilmark{4}, 
 Olivier Ilbert\altaffilmark{5}, \\Tanio Diaz-Santos\altaffilmark{6}, Johannes, Staguhn\altaffilmark{7,8}, 
Jin Koda\altaffilmark{12},  Caitlin Casey\altaffilmark{21}, David Sanders\altaffilmark{9}, \\ Bahram Mobasher\altaffilmark{10}, Nima Chartab\altaffilmark{19}, Zahra Sattari\altaffilmark{10}, Peter Capak\altaffilmark{13},  Paul Vanden Bout\altaffilmark{11}, \\ Angela Bongiorno\altaffilmark{22}, Catherine Vlahakis\altaffilmark{20},  Kartik Sheth\altaffilmark{15},  Min Yun\altaffilmark{16}, Herve Aussel\altaffilmark{17}, \\Clotilde Laigle\altaffilmark{18}, and Dan Masters\altaffilmark{2}}

\altaffiltext{}{}
\altaffiltext{1}{California Institute of Technology, MC 249-17, 1200 East California Boulevard, Pasadena, CA 91125,USA; nzs@astro.caltech.edu}
\altaffiltext{2}{IPAC, California Institute of Technology 1200 E California Boulevard, Pasadena, CA 91125,USA}
\altaffiltext{3}{Cosmic Dawn Center (DAWN), Copenhagen, Denmark, and Niels Bohr Institute, University of Copenhagen, Jagtvej 128, DK-2200 Copenhagen, Denmark}
\altaffiltext{4}{CNRS, UMR 7095, Institut d'Astrophysique de Paris, F-75014, Paris, France}
\altaffiltext{5}{Laboratoire d'Astrophysique de MarseilleLAM, Universit\'e dAix-Marseille \& CNRS, UMR7326,France}
\altaffiltext{6}{Institute of Astrophysics, Foundation for Research and Technology-Hellas (FORTH), Heraklion, GR-70013, Greece}
\altaffiltext{7}{The William H. Miller III  Department of Physics and Astronomy, Johns Hopkins University, 3400 North Charles Street, Baltimore, MD 21218, USA}
\altaffiltext{8} {Observational Cosmology Lab, Code 665, NASA Goddard Space Flight Center, Greenbelt, MD 20771, USA}
\altaffiltext{9}{Institute for Astronomy, 2680 Woodlawn Dr., University of Hawaii, Honolulu, Hawaii, 96822, USA}
\altaffiltext{10}{Department of Physics and Astronomy, University of California, Riverside, 900 University Ave, Riverside, CA 92521, USA}
\altaffiltext{11}{National Radio Astronomy Observatory, 520 Edgemont Road, Charlottesville, VA 22901, USA}
\altaffiltext{12}{Department of Physics and Astronomy, SUNY Stony Brook, Stony Brook, NY 11794-3800, USA}
\altaffiltext{13}{Spitzer Science Center, MS 314-6, California Institute of Technology, Pasadena, CA 91125, USA}
\altaffiltext{14}{North American ALMA Science Center, National Radio Astronomy Observatory, 520 Edgemont Road, Charlottesville, VA 22901, USA}
\altaffiltext{15}{NASA Headquarters, 300 Hidden Figures Way SE, Mary W. Jackson Building, Washington DC 20546}
\altaffiltext{16}{Department of Astronomy, University of Massachusetts, 710 North Pleasant Street, Amherst, MA 01003, USA}
\altaffiltext{17}{AIM Unit\'e Mixte de Recherche CEA CNRS, Universit\'e Paris VII UMR n158, Paris, France}
\altaffiltext{18}{Institut d'Astrophysique de Paris, UMR 7095, CNRS, and Sorbonne Universit\'e, 98 bis boulevard Arago, 75014 Paris France}
\altaffiltext{19}{The Observatories of the Carnegie Institution for Science, 813 Santa Barbara St., Pasadena, CA 91101, USA}
\altaffiltext{20}{National Radio Astronomy Observatory, 520 Edgemont Road, Charlottesville, VA 22903, USA}
\altaffiltext{21}{The University of Texas at Austin, 2515 Speedway Blvd Stop C1400, Austin, TX 78712, USA}
\altaffiltext{22}{INAF - Osservatorio Astronomico di Roma, Via di Frascati 33, I-00040 Monteporzio Catone, Rome, Italy}
\altaffiltext{}{}
\altaffiltext{}{}
\altaffiltext{}{}
\altaffiltext{}{}
\altaffiltext{}{}
\altaffiltext{}{}
\altaffiltext{}{}
\altaffiltext{}{}
\altaffiltext{}{}
\altaffiltext{}{}
\altaffiltext{}{}
\altaffiltext{}{}
\altaffiltext{}{}
\altaffiltext{}{}
\altaffiltext{}{}
\altaffiltext{}{}
\altaffiltext{}{}
\altaffiltext{}{}
\altaffiltext{}{}
\altaffiltext{}{}

\begin{abstract}  
ALMA observations of the long wavelength dust continuum are used to estimate the gas  masses in a sample of 708 star-forming (SF) galaxies  at z = 0.3 to 4.5. We determine the dependence of gas masses and star formation efficiencies (SFR per unit gas mass) on: redshift(z); M$_{\rm *}$; and SFR relative to the main sequence (MS).
We find that 70\% of the increase in SFRs of the MS is due to the increased gas masses at earlier epochs while 30\% 
is due to increased efficiency of SF. For galaxies above the MS this is reversed -- with 70\% of the increased SFR relative to the MS being due to elevated SFEs. Thus, the major evolution of star formation activity at early epochs is driven by increased gas masses, while the starburst activity taking galaxies above the MS is due to enhanced triggering of star formation (likely due to galactic merging). The interstellar gas peaks at z  = 2 and dominates the stellar mass down to z= 1.2. Accretion rates needed to maintain continuity of the MS evolution reach  $>100$ \msun yr$\rm ^{-1}$ at z $\rm >$ 2. The galactic gas contents are likely the driving determinant for both the rise in SF and AGN activity from $\rm z = 5$ to their peak at z = 2 and subsequent fall to lower z. We suggest that for self-gravitating clouds with supersonic turbulence, cloud collisions and the filamentary structure of the clouds regulate the star formation activity. \end{abstract} 

\date{Accepted ApJ 11/08/22}                                           

\section{Galaxy Evolution at High Redshift}

Galaxy evolution is driven by three processes: the conversion of interstellar gas into stars, the accretion of 
intergalactic gas to replenish the galactic interstellar gas reservoirs, and the interactions and merging of galaxies. The latter redistributes the 
accreted gas, promotes starburst activity, fuels active galactic nuclei (AGN) and can transform the morphology from disks 
(rotation dominated) to ellipsoidal systems. The interstellar gas plays a defining role in these processes. The 
gas being dissipative in its dynamics, centrally concentrates where it can fuel nuclear starbursts and/or AGN. 

The gas supply and its evolution at high redshifts is only loosely constrained \citep[see the reviews -- ][]{sol05,car13,tac20}. At present, only $\sim$200 galaxies at z $>$1 have been observed in 
the CO lines and most are not in the CO (1-0) line,  which is a well-calibrated mass tracer of cold molecular gas. 
 To properly understand the early evolution requires significantly larger samples of galaxies. In particular, one must probe the dependence of SFRs and star formation efficiency on the multiple independent variables: 
\begin{enumerate} 
\item the variation with redshift or cosmic time, 

\item the dependence on galaxy stellar mass ($\rm M_*$) and 

 \item the differences between the galaxies with 'normal'  SF activity on the main sequence and starburst galaxies (SB). 
\end{enumerate}

At each redshift, the majority of SF galaxies populate a relatively narrow locus in SFR versus $\rm M_*$ but this so-called Main Sequence of SF galaxies (MS) migrates dramatically to higher SFRs at higher redshift (z). 
At z $\simeq$ 2, the SB galaxies with SFRs elevated to 2 - 100 times higher levels than the MS galaxies,
constitute only 5\% of the total population of SF  galaxies \citep{rod11}. However, their contribution to 
the total SF is 8 - 14\% \citep{sar12} at z = 2 and likely larger at higher z. 
Constraining these evolutionary dependences requires both large samples and consistently accurate estimates of the redshifts, gas contents, stellar masses, and SFRs including both the 
SF sampled in the optical/UV and the dust-embedded SF probed by far-infrared observations.

The galaxies on the MS at high z would also be classified as starbursts if they were at low redshift since they have 3 - 20 times shorter gas depletion times ($\rm M_{gas} / SFR$) than typical spiral galaxies at low z. One might ask if their shorter gas depletion times are due to non-linear processes within more gas-rich galaxies (e.g. due to cloud-cloud collisions) or external mechanisms such as galactic merging.

In the work presented here, we develop and analyze a sample of 708 galaxies in the COSMOS survey field \citep{sco07} within all the ALMA archive pointings (as of June 2021) in Band 6 and 7 (1.3 mm and 850 \mum). The ALMA data are used to estimate the gas masses of each galaxy from the observed Rayleigh Jeans (RJ) dust continuum fluxes \citep{sco16} -- a technique calibrated here (see Figure \ref{empir_cal}). This technique is likely more reliable than the excited state CO emission lines which are commonly used. 

\medskip
The major questions we address are:
\begin{enumerate}
\item How do the gas contents depend on the stellar mass of the galaxies?
\item How do these gas contents evolve with cosmic time, down to the present, where they constitute typically less than 5-10\% of the galactic mass?
\item In the starburst phase, is the prodigious SF activity driven by increased gas supply or increased efficiency for converting the existing gas into stars?
\item How does the efficiency of star formation vary with cosmic epoch, stellar mass of the galaxy, 
and whether the galaxy is on the MS or undergoing a starburst?
\item How rapidly is the gas being depleted? The depletion timescale is characterized by the ratio $M_{\rm gas}$/SFR, but to date this has not 
been measured in broad samples of galaxies due to the paucity of reliable interstellar gas measurements at high redshift.
\item At present there are virtually no {\it observational} constraints on the accretion rates needed to maintain the SF at high redshifts, only theoretical
predictions. Here we provide estimates on the accretion rates needed to maintain the observed gas contents. 
\end{enumerate}

\subsection{Modifications from our Analysis in Earlier Work}

There are a number of significant differences between this work and our previous analysis \citep{sco16,sco17}. 

\begin{enumerate}
\item IR luminosities used to estimate the dust obscured SFRs are obtained from the merging of three infrared catalogs as explained in \S \ref{ir_data}. This allows reaching deeper flux levels and provides greater reliability.

\item The number of ALMA pointings is doubled as a result of there being more years in the ALMA public archive. The continuum data were limited to ALMA Bands 6 and 7. (Few pointings in the lower frequency bands (3-5) have sufficient depth to detect sources not seen in Bands 6 \& 7. )

\item The number of sources used for calibration of mass determinations from the RJ continuum is nearly doubled to 128 galaxies, although the calibration constant remains similar.

\item Our adopted MS definition now has a break in the mass dependence, curving downward above log $M_* = 10^{10.5} \msun$ \citep{lee15} rather than a single powerlaw (see Figure \ref{ms_plot}). The dependence on cosmic age (and redshift) is a power law (model \#49 from \cite{spe14}). The MS definition  thus becomes a hybrid of two terms -- one with only dependence on $\rm M_*$ and the other having only the evolutionary dependence on redshift (z).  

\item In our functional fitting for the SFEs and gas masses we use new independent variables more naturally suited than a '1+z' dependence to the evolution of the galaxy population. These separable functions for the evolution of the MS and the stellar mass dependence of the MS are now used in the functional fitting of the data. 

\end{enumerate}

In order to expedite presentation of the science results of this work, several of the detailed backgrounds to the investigation are presented in appendices: 
\begin{enumerate}
\item Appendix \ref{dust_app}: a dust heating and radiative transfer model to illustrate the use of the RJ continuum to estimate gas masses, 
\item Appendix \ref{empirical}: empirical calibration of the use of RJ flux measurements to estimate gas masses,
\item Appendix \ref{continuity}: a Continuity Principle for MS galaxy evolution, and 
\item Appendix \ref{datasets}: the input datasets and measurement procedures.
\end{enumerate}

\section{Complete Sample of ALMA-detected IR-bright Galaxies}

\medskip
\begin{figure}[ht]
\epsscale{1.15}  
\plotone{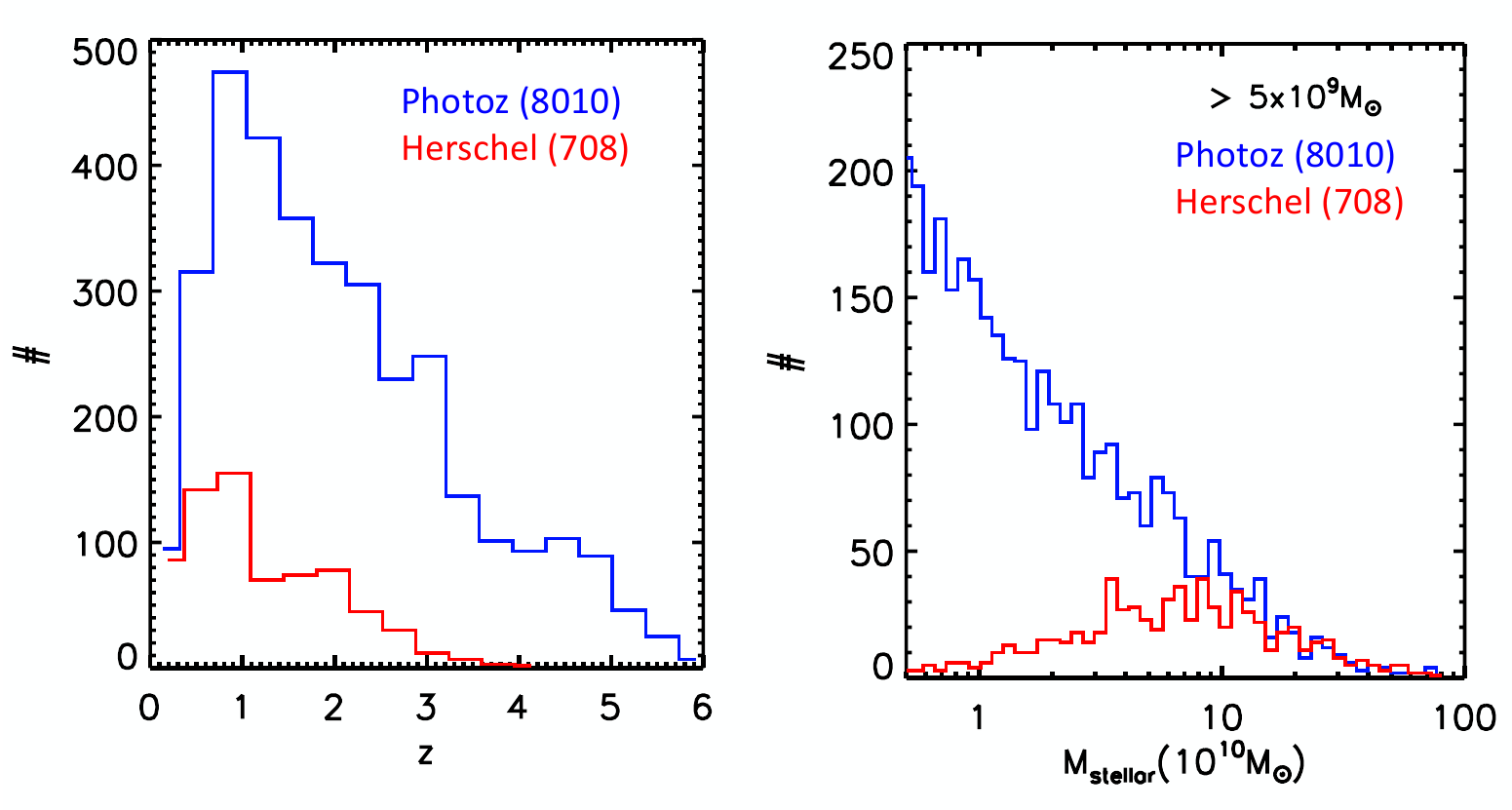}
\caption{The redshifts and stellar masses for two samples of galaxies within the ALMA pointings.
In blue, 8010 galaxies from the photometric redshift catalog \citep{wea22} sample and in red, 708 galaxies from the photometric catalog with both ALMA measurements and Herschel IR detections. The final sample for our analysis is the latter since the former didn't have either the long wavelength dust continuum needed for our gas mass estimates or the infrared for total (unobscured and obscured) SFR estimates.  }
\label{sample_m_sfr} 
\end{figure}

\begin{figure}[ht]
\epsscale{1.2}  
\plotone{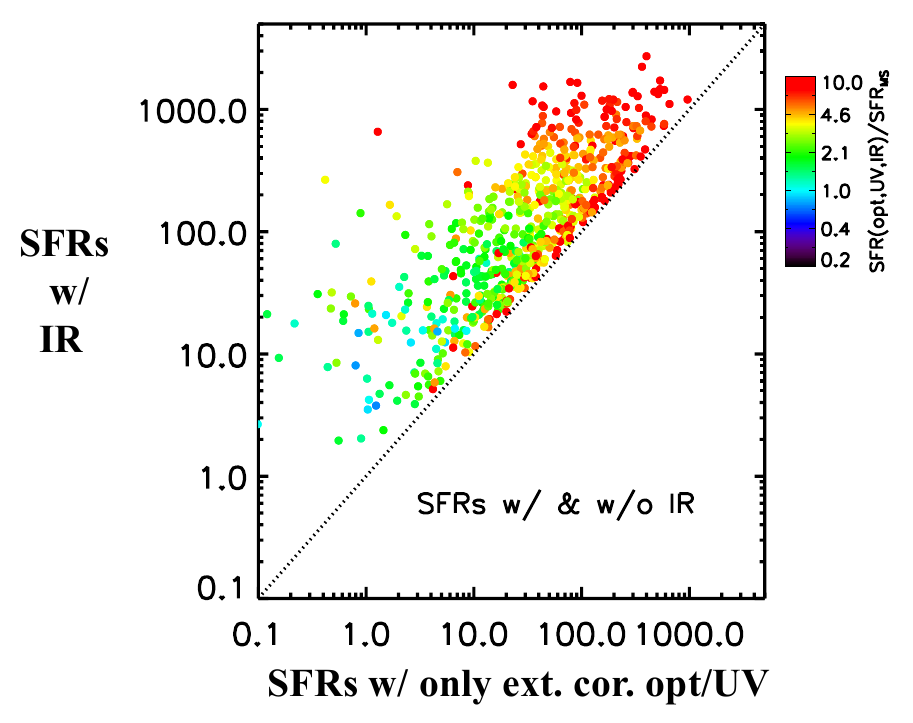}
\caption{Total SFRs for the final sample of 708 galaxies derived from the sum of both the unobscured optical/UV and the dust obscured component sampled by Herschel, are plotted against the SFR derived from opt/UV data, with extinction corrections. The colors indicate the sSFR ($\rm = SFR / M_*$) of each galaxy relative to the MS taken from the COSMOS 2020 photometric redshift catalog \citep{wea22}. {\bf This clearly demonstrates the need for direct measurements of the obscured IR component even for galaxies near the MS, since the optical/UV data alone usually underestimates the total SFR.} The numbers of galaxies at different ratios of SFR relative to the MS can be judged from numbers of points with various colors, indicating good sampling down to approximately a factor 2 above the MS SFRs.}
\label{sfr_opt_ir} 
\end{figure}


The galaxy sample used here has ALMA continuum observations in Band 6 (240 GHz) and/or Band 7 (345 GHz); they are all within the COSMOS survey field and thus have uniform quality and deep ancillary data  \citep{wea22}. The ALMA pointings are non-contiguous 
but their fields of view (FOV), totaling 102.9 arcmin$^2$, include 708 galaxies with measured far-infrared fluxes from the Spitzer and Herschel Observatories. This sample also has calibrations that are uniform across the full sample without the
need for zero-point corrections.Their redshift and stellar mass distributions are shown in Fig. \ref{sample_m_sfr}. All of the 
Herschel sources within the ALMA pointings are detected by ALMA. There are of course sampling biases in the different areas of the parameter space but the multivariable fitting described below should continuously join those areas.

\begin{figure}[ht]
\epsscale{1.2}  
\plotone{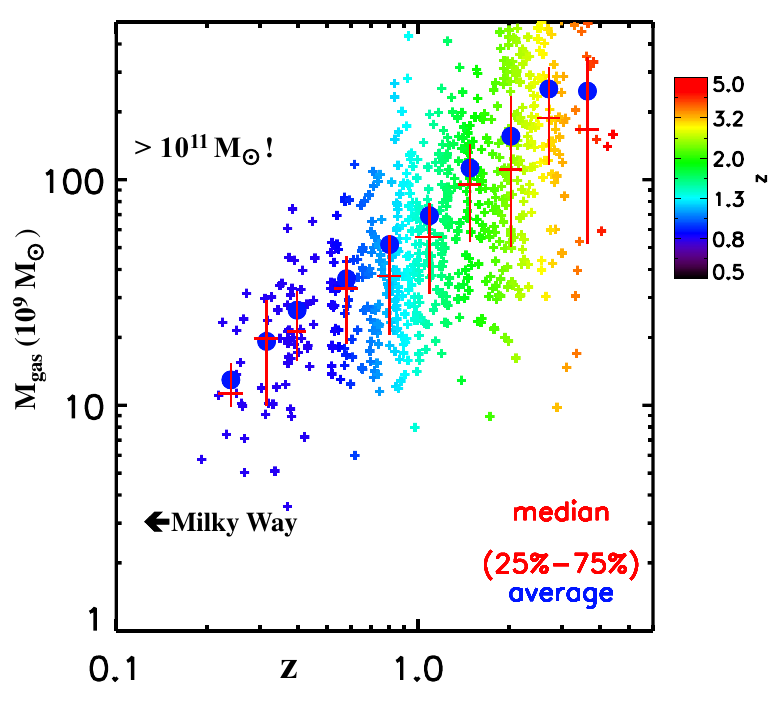}
\caption{The estimated gas contents ($M_{\rm gas}$) are shown for 708 in the ALMA Herschel sample and are plotted as a function or redshift.   
The measured gas masses and SFEs are shown binned by redshift with the median for each bin as a horizontal line and the average as the blue dot. The vertical line indicates the range between the 25 and 75 percentiles of each bin.  The arrow indicates the molecular 
gas mass of the Milky Way for reference as $\sim 3\times10^9$ \msun. Many high redshift galaxies have gas contents over $10^{11}$ \msun.  [For clarity, in Figures (\ref{z_dep} - \ref{ssfr_dep}) below we do not show the individual points .]}
\label{sfr_ism} 
\end{figure}

The dusty SF activity is, in virtually all cases, dominant (5-10 times) over the unobscured SF probed in the optical/UV.
This is clearly demonstrated in Fig. \ref{sfr_opt_ir}. The use of SFRs based on opt/UV data alone with extinction corrections derived from the opt/UV data, can result in an order of magnitude under-estimation of the total SFR, even for 
galaxies close to the MS. \footnote[1]{This may also suggest that the apparent tightness of the MS seen in some OPT/UV-only studies is in part due to not properly accounting for the dust obscured SFR component. Clearly, the estimation of the overall extinction from opt/UV data alone will only be sampling the less
 obscured SFR on the outskirts of the SF regions, not the full line of sight. Even in low redshift galaxies having much lower gas masses, a typical Giant Molecular Cloud (GMC) has $A_V = 20 $ mag implying a visual extinction of a factor $10^8$ and higher in the UV.}

\begin{figure*}[ht]
\epsscale{0.8}  
\plotone{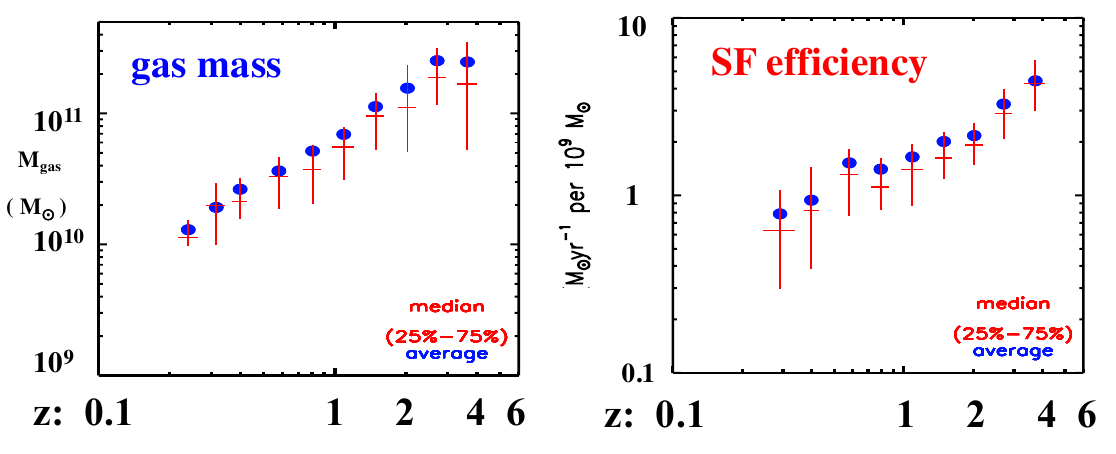}
\caption{The measured gas masses and SFEs are shown binned by redshift with the median for each bin as a horizontal line and the average as the blue dot. The vertical line indicates the range between the 25 and 75 percentiles of each bin. The data suggest that the dominant change with redshift is in the gas masses (a factor of $\sim 20$ increase) while the SFE is approximately half as much in the power law increase (a factor 4 increase). Thus $\sim$2/3 of the evolution is due to increasing gas contents and 1/3 is increasing efficiency.}  
\label{z_dep} 
\end{figure*}

\begin{figure*}[ht]
\epsscale{0.8}  
\plotone{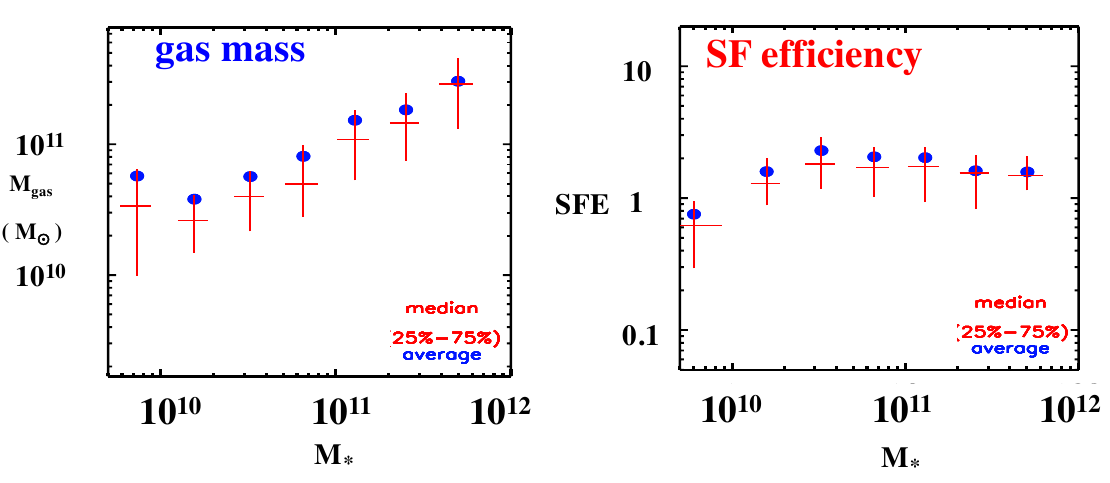}
\caption{The measured gas masses and SFEs are shown by stellar mass $(M_*)$ with the median for each bin as a horizontal line and the average as the blue dot. The vertical line indicates the range between the 25 and 75 percentiles of each bin. These data show increasing gas masses for higher $(M_*)$ galaxies but with on a dependence $(M_*^{0.67} )$, while the SFE is virtually independent of $(M_*)$.} 
\label{mstar_dep} 
\end{figure*}

\begin{figure*}[ht]
\epsscale{0.8}  
\plotone{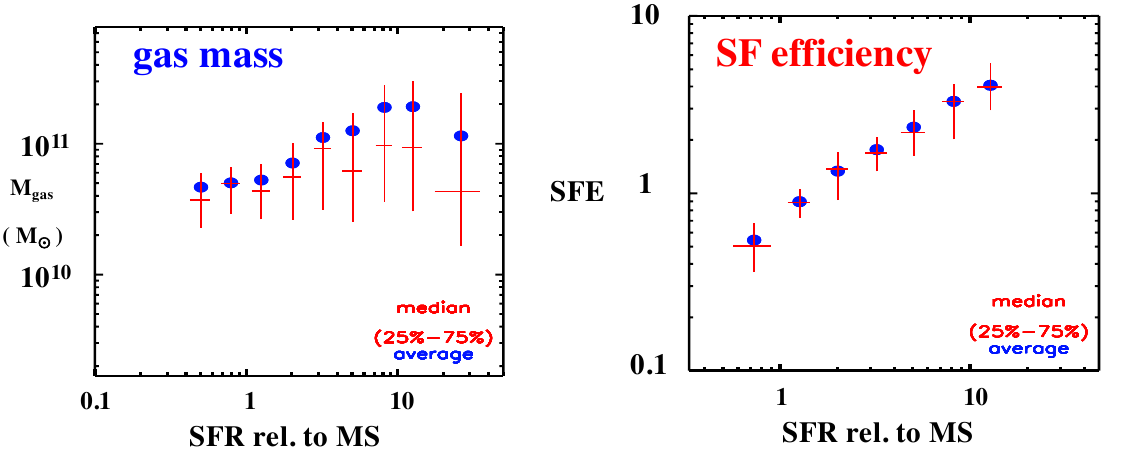}
\caption{The measured gas masses and SFEs are shown binned by specific star formation rate relative to the MS with the median for each bin as a horizontal line and the average as the blue dot. The vertical line indicates the range between the 25 and 75 percentiles of each bin. The dominant trend as one goes above the MS is an increased rate of forming stars per unit gas mass, with a less systematic variation of gas mass.} 
\label{ssfr_dep} 
\end{figure*}

\subsection{Distributions of M$_{gas}$ and SFE}\label{ism_fitting}

The ALMA fluxes were translated into gas masses using Eq. \ref{dust_eq}. 
These gas masses are shown in Figure \ref{sfr_ism} for the entire sample of 708 detected sources. As clearly seen in the redshift color coding of the 
points, there must be strong dependence of the gas masses on z. Yet this can not be the entire explanation of the scatter since the high redshift galaxies 
with a large range of gas masses exhibit similarly high SFRs, implying that the rate of SF per unit gas mass (the SFE) is also varying (see Figure \ref{z_dep}-Right).

Figures \ref{z_dep} - \ref{ssfr_dep} show the trends in the gas contents and star formation efficiency with cosmic time, stellar mass, and whether 
the galaxies are on the MS or have elevated SFRs and are in the starburst region. 
However, these plots do not adequately take account of the joint, simultaneous dependences on the independent parameters (z, $M_*$ and MS versus SB galaxies). [The latter is parameterized by SB below.]  We fit for all of these dependencies simultaneously in Section \ref{fits} below.

\subsection{Simultaneous Fitting for the Joint Dependences of M$_{\rm gas}$, SFR and SFE}\label{fits}

\begin{figure}[]
\epsscale{1.}  
\plotone{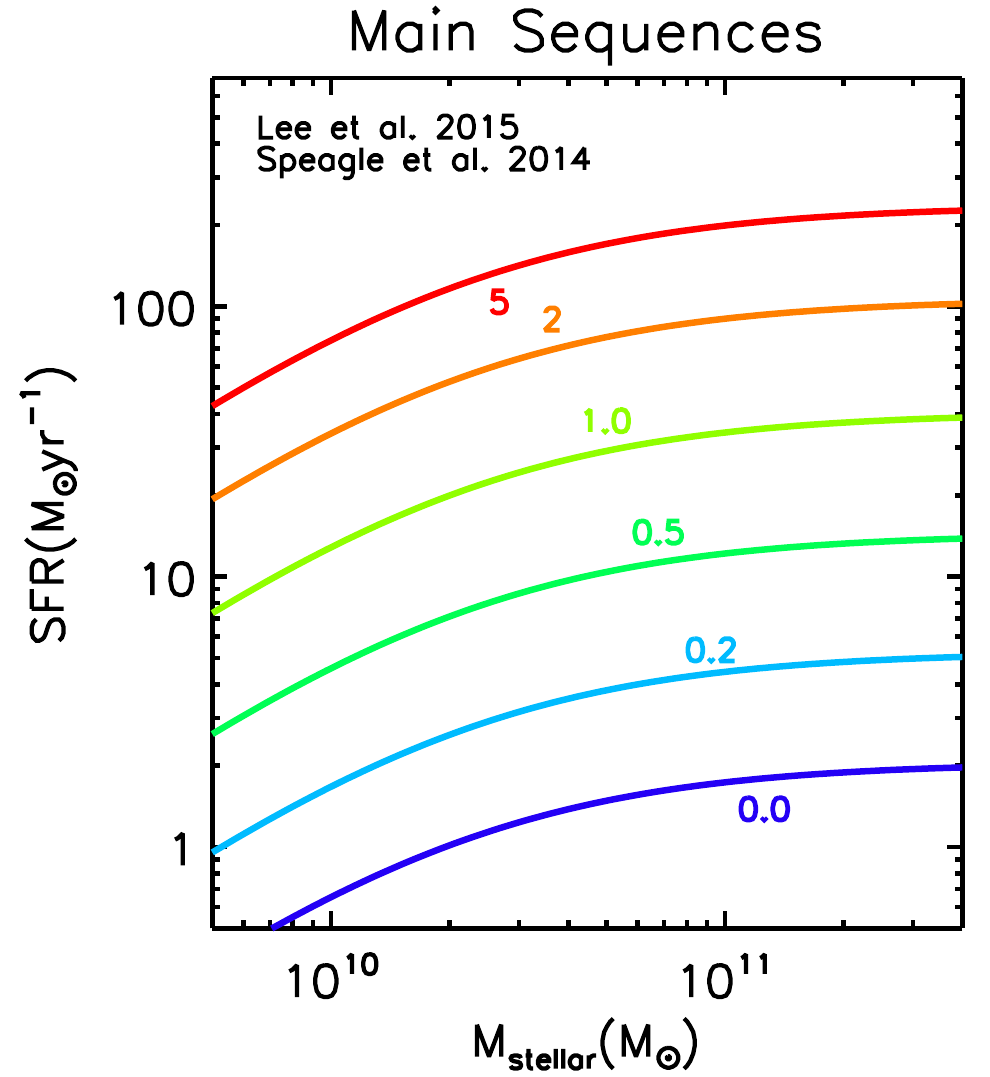}
\caption{The adopted star formation main sequence (MS) is a hybrid product of the evolution with cosmic time (or z) from \citep[][fit \# 49]{spe14} and as a function of stellar mass  the z = 1.2 MS of \cite{lee15}. Two of the independent variables used in the fitting are ${\bf SFR_{MS} (z)}$ and ${\bf SFR_{MS} (M_*)}$ representing these two
terms: the evolutionary dependence of the MS with z and the shape of the MS as a function of $M_*$. The third MS variable is {\bf SB}, characterizing departure from the MS
into the starburst region. This is the ratio of each galaxy's total SFR to its SFR if it were on the MS (at its stellar mass and redshift). 
}
\label{ms_plot} 
\end{figure}

Simultaneous fits were done using two techniques: Monte Carlo Markov Chain (MCMC, MLINMIX\_ERR, \citep{kel07}) 
and Levenberg-Marquardt least squares fitting (lm\_fit.pro in IDL).
For each fitting, the terms for which power-law coefficients were obtained included:

\begin{enumerate}
\item ${\bf SFR_{MS} (z)}$, representing the temporal evolution of the MS as a function of z,
\item ${\bf SFR_{MS} (M_*)}$, encapsulating the mass dependent shape of the MS (shown in Fig. \ref{ms_plot} by the z = 0 MS curve),
\item {\bf SB} to quantify dependences for starburst galaxies above the MS. Specifically, SB is equal to the ratio of a 
galaxy's total SFR (=$SFR_{opt/UV} + SFR_{IR}$) to that of a galaxy on the MS at the same redshift and stellar mass,  and
\item ${\bf M_*}$, to capture any simple dependence on the stellar mass of the galaxies.
\end{enumerate}  

Two of these terms represent variations relative to the MS: 1) with respect to cosmic age ($SFR_{MS} (z)$) and 2) with respect to SFR as a function of stellar mass along the MS (${SFR_{MS} (M_*)}$). As noted above, SB measures the total SFR relative to the MS.
The use of separate functions for the evolution in time and mass relative to the MS and elevation above the MS allows one to
probe the evolutionary dependence and the difference between the SB population and the MS galaxies. The steepness of the power-law coefficients for each term enables judgement of the relative importance to
changes in the gas contents and the star formation efficiency. 

Figure \ref{ms_plot} shows the MS loci in the $\rm M_*$ and SFR plane for redshifts z = 0 -- 5. The shape of the MS as a function of mass is taken from the z = 1.2 MS of \cite{lee15} and the evolution as a function of cosmic age (i.e. z) is from \cite{spe14} fit \# 49, normalized to a fiducial mass of log M$_{\rm *} = 10.5$ \msun. The same mass dependence of the MS is used for all z. 

\medskip

Both the MCMC and LM solutions gave similar coefficients within
the uncertainties (2-10\% of the values of the coefficients). The primary advantage of the MCMC fitting is that it fully probes the 
parameter space, while the LM fitting takes account of uncertainties in the dependent variables and is much faster. The fitting results are listed in Table \ref {equations1}; the coefficients listed in Table \ref{equations1} are from the LM fit. Additional relations for the gas depletion time and gas mass fraction are easily derived from those equations so they are not listed here. The last relation in the Table for the net gas accretion rate is derived in \S \ref{accretion}.

\begin{deluxetable*}{lll}[h]

\large
  \tablenote {$\rm M_{\rm * 10}$ is $\rm M_{stellar}$ in units of $\rm 10^{10}\msun$. The fitting was done using both Monte Carlo Markov Chain and Levenberg-Marquardt techniques as detailed in the text, yielding similar exponents for the fit terms (within $\sim 5\%$). Eq. 1-3 are from the LM fitting. Equation 3 is obtained from the product of the first two equations.}
\tablecaption{\bf{M$_{\rm {\bf gas}}$, SFR and SFE  FITTING }}

\tablehead{\colhead{} & \colhead{Eq. \#}    }
\startdata 
\\  
\boldmath $ \rm M_{\rm gas} = \rm 2.68  \times 10^9~\msun    \times SFR_{MS}(z)^{0.70}  \times SFR_{MS}(M_*)^{-0.56}   \times \left( {\rm M_{\rm * 10}}\right)^{0.45}  \times (\rm SB)^{0.29} $  & {\bf 1}  \label{ism_fit} \\
\\ 
\\
 \boldmath $ \rm SFE  = 
 \rm 0.25 \times SFR_{MS}(z)^{0.33} \times SFR_{MS}(M_*)^{1.81}   \times \left( {\rm M_{\rm * 10}}\right)^{-0.60}  \times  \left( \rm SB \right) ^{0.71}   $ & {\bf 2}  \label{sfe_fit}  \\
\boldmath ~~~~~~~~~~~$\rm (\msun yr^{-1} per~ 10^9 \msun of~ gas) $\\ 
\\
\\
 \boldmath $ \rm SFR  = \rm 0.66~\msun~yr^{-1} \times SFR_{MS}(z)^{1.03}  \times SFR_{MS}(M_*)^{1.25}    \times \left( {\rm M_{\rm * 10}}\right)^{-0.15}  \times \left( \rm SB\right) ^{1.0} $  & {\bf 3}  \label{sfr_fit}  \\
 \\
\\
 \boldmath $ \rm ACC  = \rm 0.22~\msun~yr^{-1} \times SFR_{MS}(z)^{1.70}   \times \left[  {\rm M_{\rm * 10}}^{0.62} -0.31 \left( {\rm  M_{\rm * 10} }\right)^{1.13} \right] $ & {\bf 4}  \label{acc_fit} 
 \\
 \enddata \label{equations1}

 \end{deluxetable*}
 
 \begin{figure*}[ht]
\epsscale{1.2}  
\plotone{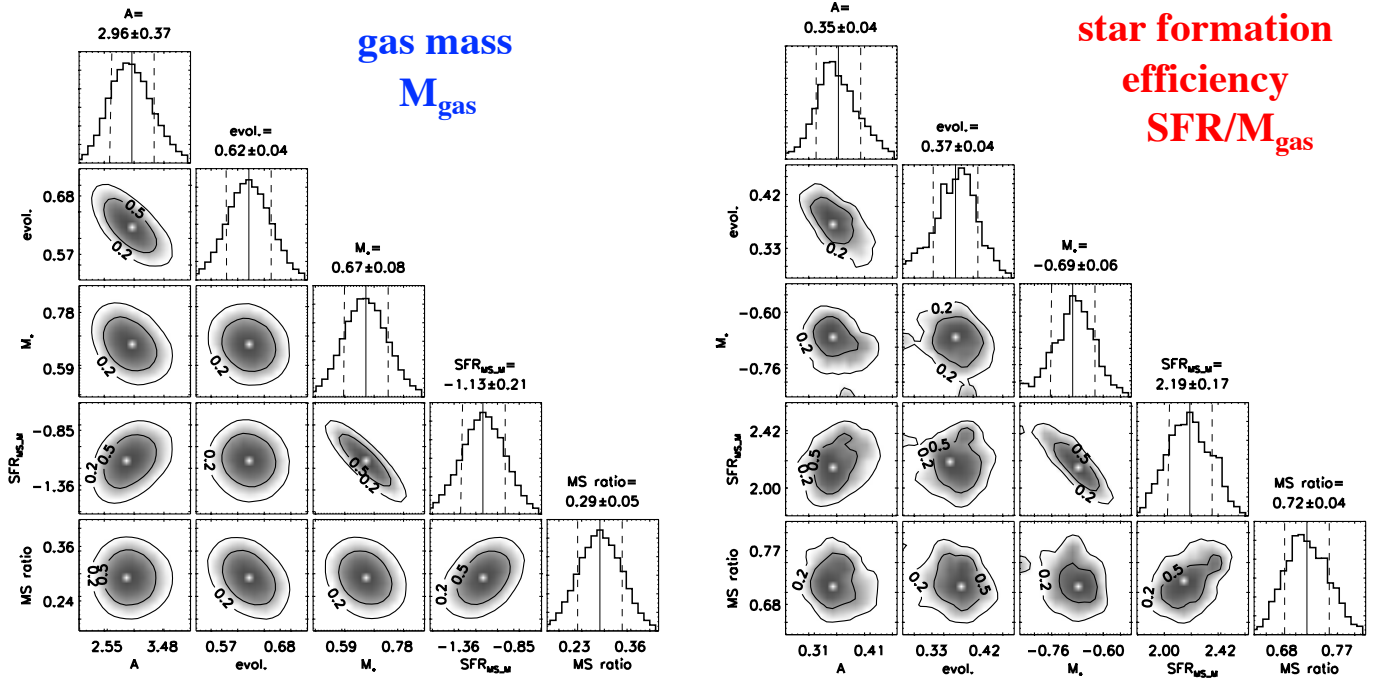}
\caption{Covariance plots from the MCMC fitting exhibit well behaved single peaked probability distributions.}  
\label{ssfr_dep_mcmc} 
\end{figure*}

Equations 1 and 2 in Table \ref{equations1} quantify major results regarding the gas contents and star formation efficiency in high redshift galaxies relative to those at low redshift galaxies, and starbursts versus MS galaxies:
\begin{enumerate}
\item The gas masses clearly increase at higher z, varying as $\rm SFR_{MS}(z)^{0.70}$ (i.e. 70\% of overall temporal dependence of the MS SFR). The remaining 30\% of the increase in MS SFR is due to increased SFE at higher redshifts (based on the exponents of the $SFR_{MS}(z)$ terms in Equations 1 and 2).
\item The reverse is true for the starburst galaxies above the MS -- the dominant driver for the activity is an increased SF efficiency -- the gas contents increase only as the 0.29 power of SB while the SFE increases as the 0.71 power of SB.  
Thus the starburst activity is largely driven by increased efficiency in forming stars per unit mass of gas and only $\sim$30\% due to increased gas masses. 
\item From the terms involving $M_*$, it is also apparent that the higher stellar mass galaxies have higher gas contents, but not in proportion to $M_*$ (i.e. the gas mass fraction decreases in the highest mass galaxies). 
\end{enumerate}

The first conclusion clearly implies the SF efficiency must increase at high redshift (as discussed below). The second conclusion indicates that the galaxies above the MS have higher gas contents, but not in proportion to their elevated SFRs. This confirms that these galaxies are undergoing bursts of activity, rather than long term elevated SFRs. 
The galaxies with higher stellar mass likely use up their fuel at earlier epochs, and have lower specific accretion rates (see Section \ref{accretion}) than the low mass galaxies. This is a new aspect of `downsizing' in the cosmic evolution of galaxies. Perhaps many of the high mass galaxies undergo  
a {\it last} fatal starburst which rapidly exhausts their gas supplies.

\subsection{Increased SFR versus higher SFE}

In fitting for the SFR dependencies, we intend to clearly distinguish the 
obvious intuition that when there is more gas there will be both more SF and a higher \emph{efficiency} for converting the 
gas to stars. Thus, in solving for the SFE we impose a fixed, linear dependence of the SFR on M$_{\rm gas}$. We are then effectively fitting for the {\emph star formation efficiencies} ($\rm SFR/M_{gas}$) for star formation per unit gas mass as a function of 
z, SB and M$_{\rm *}$. This isolates the SFE variation with redshift, SB  
and M$_{\rm *}$ from the variation in M$_{\rm gas}$ with the same three parameters. 

\medskip
\begin{figure}[ht]
\epsscale{1}  
\plotone{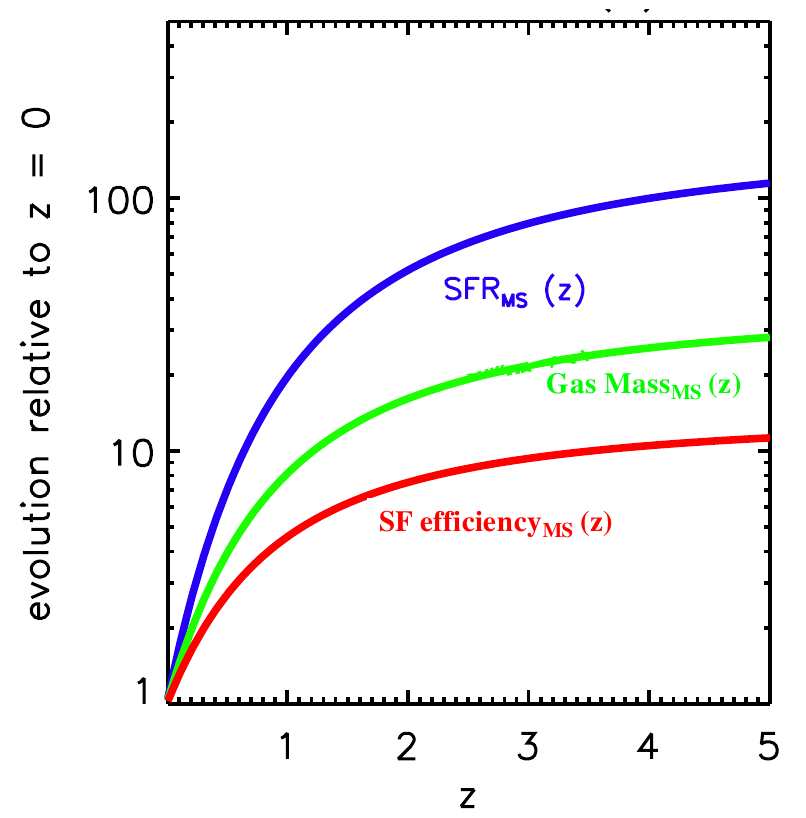}
\caption{The evolutionary dependence of the SFRs (blue), the gas masses (green) and the SF efficiency (red) per unit mass of gas
on the MS at a characteristic stellar mass of $5\times10^{10}$ \msun.}  
\label{evolutionary_dependence} 
\end{figure}

The lack of strong dependence of the SFE on galaxy mass ($\rm M_*^{0.6}$ in Eq. 2) is reasonable. If SF gas at high redshift is in \emph{self-gravitating} GMCs as at low z; the very local gravity (environment) near the GMCs may influence the internal SF but the GMC wlll not 'care' that it is in the distant potential well of a more or less massive galaxy. 

Figure \ref{evolutionary_dependence} shows the relative evolutionary dependencies of the SFRs, the gas masses and the SF efficiency per unit 
mass of gas normalized to unity at z = 0.
The fundamental conclusion is that \emph{the elevated rates of SF activity at both 
high redshift and above the MS are due to both increased gas contents and increased efficiencies 
for converting the gas to stars.}

 \begin{figure}[ht]
\epsscale{1.}  
\plotone{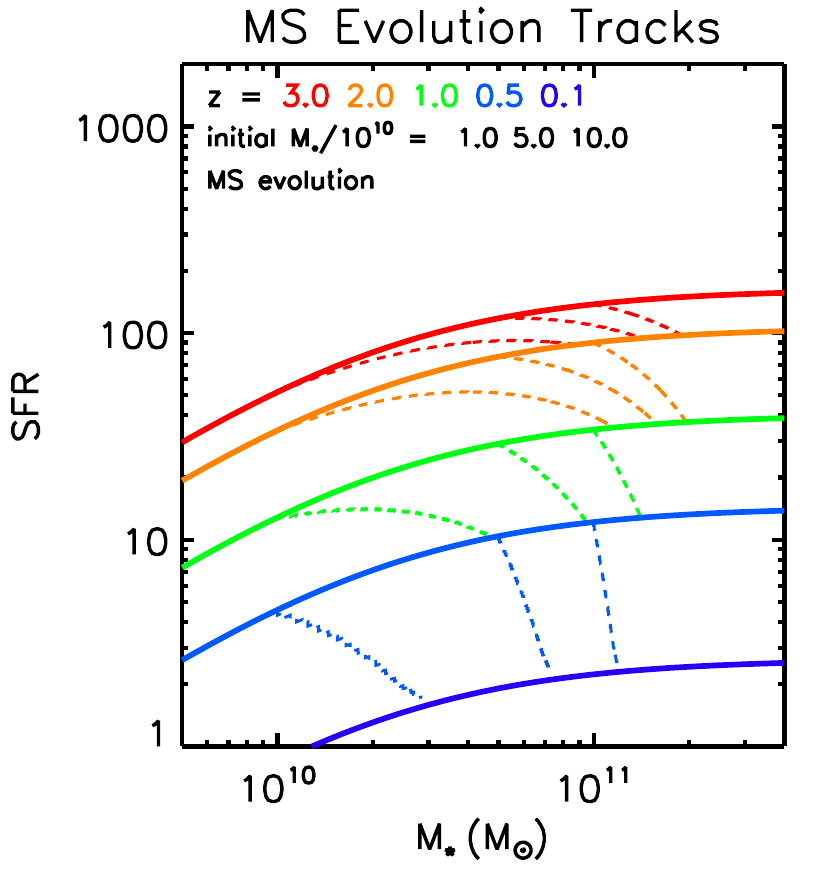}
\caption{Evolutionary tracks for MS galaxies are shown for 5 redshifts indicating that the primary transformation at high z
is toward higher $M_*$ whereas at low z the major evolution is downward in SFR.  The evolution is followed in \S \ref{accretion} to
estimate the required accretion rates needed to match the gas contents estimated above. NEW FIGURE 10/21/22 }
\label{evolution_tracks} 
\end{figure}

 \begin{figure*}[ht]
\epsscale{1.2}  
\plotone{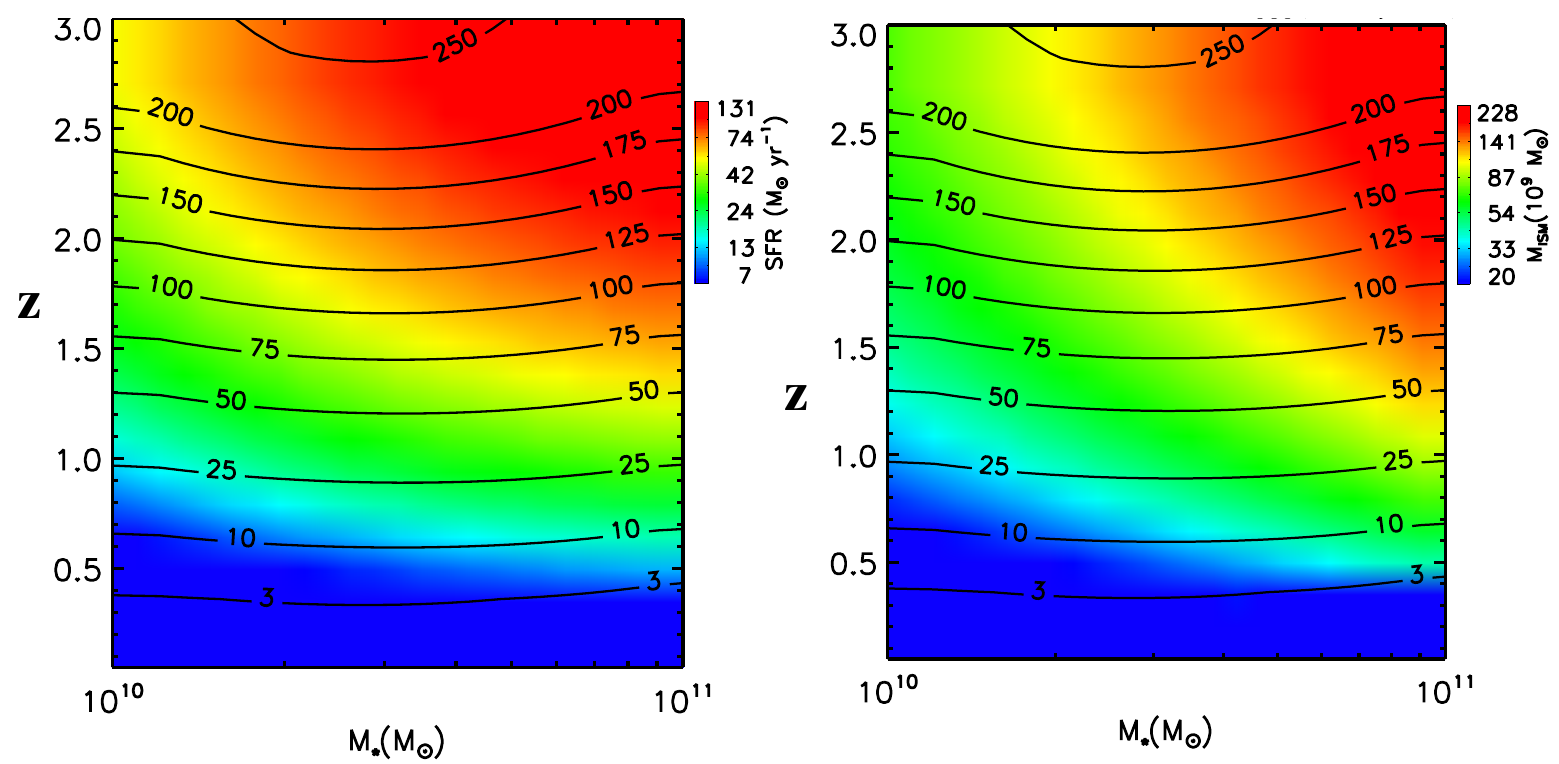}
\caption{The calculated gas accretion rates (contours) are compared with the SFRs (color in Left panel) and gas contents (color in Right panel).}
\label{accretion_rate_fig} 
\end{figure*}

\section{Gas Accretion Rates}\label{accretion}

Using the MS Continuity Principle (Appendix \ref{continuity}) and the gas contents obtained from Equation \ref{ism_fit} for the MS ( i.e. $\rm SB \equiv SFR / SFR_{MS} = 1$), we can now derive the net accretion rates of MS galaxies required to maintain the MS evolutionary tracks (see Fig. \ref{evolution_tracks}). This accretion may be purely gaseous or via minor mergers. 
This analysis is based on the simple logic that we now have estimates of the gas contents which the MS galaxies should have at each redshift as they evolve to lower redshift. And if through their SF activity the galaxies use up their gas at too great a rate and arrive at a later MS curve with a lower $\rm M_{\rm gas}$ than Eq. \ref{ism_fit}, then the difference must, on average, be made up by accretion. 
Along each evolutionary track (dashed lines in Figure \ref{evolution_tracks}), the rate balance must be given by:
\begin{eqnarray}  
{\rm dM_{\rm gas} \over{dt}} = \rm  \dot{\rm M}_{acc} - (1-\rm f_{mass~return}) \times SFR, \label{mdot_eq}
 \end{eqnarray} 

\noindent assuming that major merging events are rare. $\rm f_{mass~return}$ is the fraction of stellar mass returned to the gas through stellar mass-loss, taken to be 0.3 here \citep{lei11}. Since these paths are following the galaxies in a Lagrangian fashion, the time derivatives of a mass component M must be taken along the evolutionary track and
\begin{eqnarray}  
{\rm d M\over{dt}} &=& {\rm d M\over{dz}} {\rm dz \over{dt}} + {\rm d M\over{dM_{\rm *}}} {\rm dM_{\rm *} \over{dt}} \nonumber \\
&=& {\rm d M\over{dz}} {\rm dz \over{dt}} + \rm SFR {\rm d M\over{dM_{\rm *}}}  \label{continuity_equation}.
 \end{eqnarray} 
 
  Figure \ref{accretion_rate_fig} shows the required gas accretion rates (contours) from Equation 4 in Table \ref{equations1}. The background colors indicates the SFRs (Left panel) and the gas contents (Right panel). The required rates are $\sim100$ \msun yr$^{-1}$ at z $\sim$ 2. The combination of two separate stellar mass terms in Eq. 4 is required to match the curvatures shown in Figure \ref{accretion_rate_fig}; they don't have an obvious physical justification. The first term dominates at low mass and the second at higher masses.

 Two important points to emphasize are: 1) these accretion rates should be viewed as \emph{net rates} (that is the accretion from the halos 
 and satellite galaxies minus any outflow rate from SF or AGN feedback) and 2) these rates 
 refer only to the MS galaxies where evolutionary continuity is a valid assumption. 
  
 The derived accretion rates are required in order to maintain the SF in the early universe galaxies. Even though the existing gas 
 contents are enormous compared to present day galaxies, the observed SFRs will deplete this gas within $\sim5\times10^8$ yrs; this is short compared to MS evolutionary timescales. The large accretion rates are comparable to the SFRs. 
 The higher SF efficiencies deduced for the high-z galaxies and for galaxies above the MS may be dynamically driven by the infalling gas and galactic merging.  These processes will shock compress the galaxy disk gas, since the induced velocities are likely larger than the internal supersonic turbulence within the clouds.

  It is worth noting that although one might think that the accretion rates could have been readily obtained simply from the evolution of 
the MS SFRs, this is not the case. One needs the mapping of $\rm M_{\rm gas}$ and its change with time in order to estimate the first term on the 
right of Equation \ref{mdot_eq}. Figure \ref{accretion_rate_fig} shows the relative evolution of each of the major rate functions 
over cosmic time and stellar mass. Comparing the proximity of the curves as a function of redshift, one can see modest 
differential change in the accretion rates and SFRs  as a function of redshift and stellar mass.

\begin{figure}[ht]
\epsscale{1.}  
\plotone{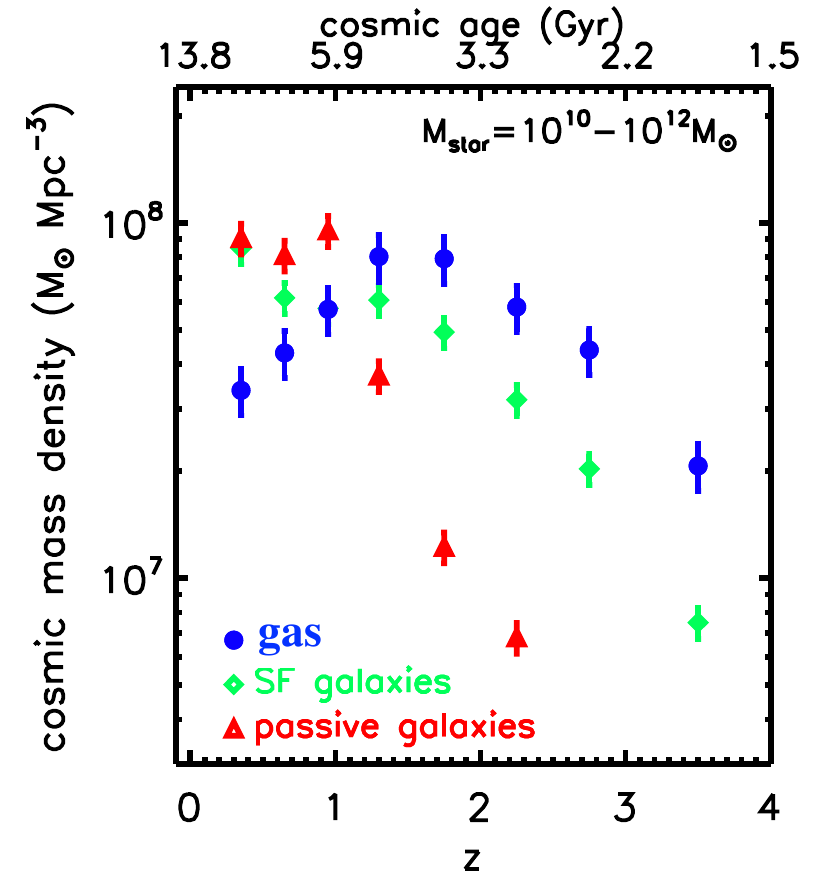}
\caption{The cosmic evolution of interstellar gas and stellar mass densities in the universe are shown for galaxies with stellar masses $\rm M_{\rm *} = 10^{10}$ to $10^{12}$ \msun.
 The galaxy stellar mass functions from \cite{ilb13} were used to calculate the gas masses using Equation \ref{ism_fit}. Uncertainties in the stellar mass densities are typically $\pm10$\% for this range of $\rm M_{\rm *}$ \citep[see][Figure 8]{ilb13}; uncertainties in the gas mass density also include an uncertainty of $\pm10$\% in the gas masses when averaged over the population. (This does not include uncertainty in the calibration of the dust-based mass estimations.)}.
\label{lilly_madau}  
\end{figure}

\begin{figure}[ht]
\epsscale{1.}  
\plotone{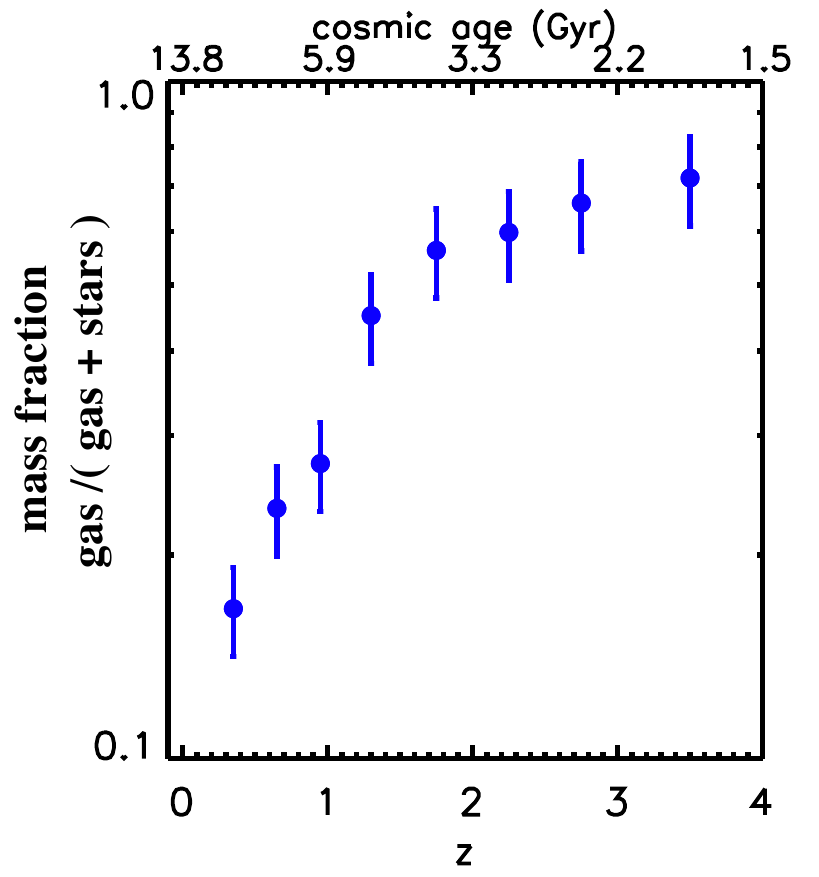}
\caption{The mass fraction of gas is shown for galaxies with stellar masses $\rm M_{\rm *} = 10^{10}$ to $10^{12}$ \msun.}
\label{mass_frac}  
\end{figure}

\section{SB versus MS Galaxy Evolution}\label{sb_ms}
 
 One might ask what are reasonable accretion rates to adopt for the SB galaxies above the MS, since they are not 
 necessarily obeying the continuity assumption? Here there appear two possibilities: either the accretion rate is similar to
 the MS galaxy at the same stellar mass, or if the elevation above the MS was a consequence of galactic merging, one might assume a rate 
 equal to twice that of an individual galaxy with half the stellar mass. The latter assumption would imply $\sqrt{\rm 2}$ higher accretion rate, thus 
 being consistent with the higher gas masses of the galaxies above the MS. In this case, the higher SFRs will be maintained longer 
 than the simple depletion time it takes to reduce the pre-existing gas mass back down to that of a MS galaxy. The same $\sqrt{\rm 2}$ factor of increase in the gas mass 
 will arise from the merging of the pre-existing gas masses of two galaxies of half the observed mass. This follows from the dependence of M$_{\rm gas}$ on stellar mass, varying only as M$_{\rm *}^{0.30}$, rather than linearly. \emph {Thus, the notion that the SB galaxies 
 are the result of galaxies merging is favored. }

\section{Cosmic Evolution of Gas and Stellar Mass}\label{cosmic}

Using the mass functions (MF) of SF and passive galaxies \citep{ilb13}, we estimate the total cosmic mass density of gas as a function of redshift using 
Equation \ref{ism_fit}. (This is the equivalent of the Lilly-Madau plot for the SFR density as a function of redshift.) We do this for the redshift range z = 0 to 4
and $M_{\rm *} = 10^{10}$ to $10^{12}$ \msun, a modest extrapolation of the ranges covered in the data presented here. Figure \ref{lilly_madau} shows 
the derived cosmic mass densities of stars (SF and passive galaxies) and gas as a function of redshift.  We applied Equation \ref{ism_fit} only to the SF galaxies and did not include any contribution from the passive galaxy population; to include the SB population we multiplied the gas mass of the normal SF population by a factor of 1.1.  If the galaxy distribution is integrated down to stellar mass equal to $10^9$ \Msol, the stellar mass (and presumably the gas masses) are increased by 10 to 20\% \citep{ilb13}.  

The evolution of the gas mass density shown in Figure \ref{lilly_madau} is similar in magnitude to the theoretical predictions based on semi-analytic models by \cite{obr09,lag11,sar13} (see Figure 12 in \cite{car13}). However, all of their estimations exhibit a more constant density at z $>$ 1. The empirically based, prescriptive predictions of \cite{pop15} exhibit closer agreement with the evolution found by us; they predict a peak in the  
gas at z $\simeq$ 1.8 and a fall-off at higher and lower redshift. (All of those previous estimations have much larger uncertainties.)

The gas mass fractions computed for galaxies with $M_{\rm *} = 10^{10}$ to $10^{12}$ \msun~ are shown in Figure \ref{mass_frac}. The gas is dominant over the stellar mass down to z $\simeq 1.5$.  At z = 3 to 4 the gas mass fractions 
get up to $\sim 80$\% when averaged over the galaxy population. Thus, gas contents which peak at z $\simeq$ 2, are likely responsible for the peak in SF and AGN at that epoch, since the latter are dependent on the gas for fueling.  At z = 4 down to 2, the buildup in the gas density is almost identical to that of the cosmic SFRD. (Note that the gas density point at z $\sim 0.3$ is uncertain since it relies on extrapolation of 
Equation \ref{ism_fit} to low M$_{\rm *}$ and low z, where there exist relatively few galaxies in our sample.) 

\section{Still ... a Lot to be Learned from Local Galaxies}\label{local} 

\begin{figure} [ht]
\epsscale{1.2} 
\plotone{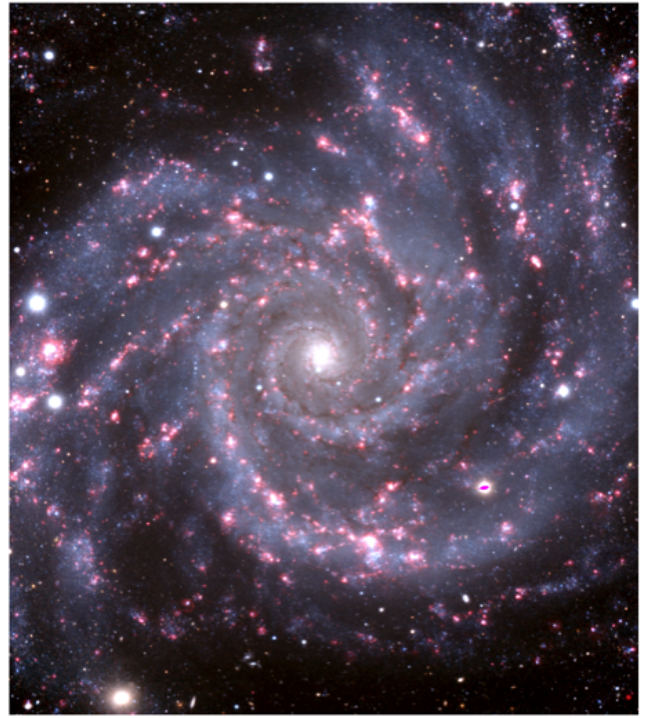} 
\caption{Optical image of M 74 made with Subaru Suprime-Cam data (Koda \etal)}
\label{m74} 
\end{figure}

All of the foregoing has focused on the global properties of galaxies at high redshift. The internal distributions of gas, SF and stars are
 critical to developing a physical understanding of the nature of the gas clouds and the SF within them. To inspire this discussion,   
a visible wavelength, multiband image of M74 is provided in Fig. \ref{m74}. It shows the dramatic organization of the star cluster formation along the spiral arms. The thinness of such galactic disks is due to the dissipative nature of the SF GMCs which damp the gas motions perpendicular to the disk. 
In high z galaxies, which are accreting gas at high rates and have large energy and momentum input from both the accretion and the active SF, the disk is also likely to be more irregular in structure \citep[e.g. ][]{for11}. 

In nearby galaxies, essentially all of the SF takes place within the GMCs, not in the HI. In the Milky Way disk the GMCs have extraordinary properties -- some of which are yet to be understood. The half mass point for the GMC population is at $\rm 2\times10^5$ \msun, diameter 40 pc and mean density 
$\rm n_{H2} = 300$ cm$^{-3}$. Although the gas thermal temperature is $\sim 10 - 20$K (sound speed $\sim \rm 0.1 -- 0.2 ~km s^{-1}$), the molecular emission line widths are typically 5 km s$^{-1}$, indicating 
Mach $\gtrsim 10$ supersonic turbulence. The magnitude of the turbulence is marginally contained by the self-gravity of the molecular gas given the sizes, masses 
and velocity dispersions. Hence, they are self-gravitating and do not fly apart within the cloud crossing timescale of a few Myr (see Scoville \& Sanders 1987 for a summary of GMC properties). The source of the 
turbulent pressure support is not well constrained. Although we routinely visualize the clouds as spherical for ease of analysis, that is certainly not the case --
they are often filamentary -- indicating that they are only marginally gravitationally bound. Both the filamentary structure and the supersonic turbulent support may reflect the importance of internal magnetic fields. 

In the central 300 pc of the Galaxy, there are a number of 
molecular clouds with more extreme properties -- $10^6$ \msun and velocity dispersions $10 - 20$ km s$^{-1}$. We have found in the high z galaxies total molecular gas masses 10 - 100 times that of the Galaxy. We expect that the gas clouds there will be much more massive than those in the local galaxies -- this intuition guided our modeling of the IR radiative transfer in clouds of $10^{6 - 8}$ \msun (see Appendix 1). 

\smallskip
\underline{For how long do the GMCs and their $\rm H_2$ survive?} One might think that given the marginal self gravity of the GMCs and the energy releases from SFR, that 
the GMCs last no more than a few cloud crossing times (i.e. $10^{6-7} $ yrs). However, there is a very simple argument suggesting that they last more than $10^8$ yrs. In the inner disks of local SF galaxies, it is always the case that the molecular gas dominates the atomic gas in mass. Within each annulus around the center of the galaxy there must be conservation of the mass flux from $\rm H_2$ to HI+HII and vice versa (since the SF consumption is relatively small for each orbit) : 

\begin{eqnarray}  
 {\rm <\tau_{H2}> } = <\tau_{HI}> \times  {{M_{H2}}\over{M_{HI}}}
\end{eqnarray}

\noindent where $<\tau>$ is the mass-weighted average timescale of each phase, and HII is neglected since it is generally a minor mass component. Since the typical timescale for the HI is at least $\sim10^8$ yrs to get to the next spiral arm, and the H$_2$/HI mass ratio is 
typically greater than 10:1 in the interior regions of the galactic disks, the lifetime of $\rm H_2$ should be 10$^{8-9}$ yrs. An alternative way to phrase this point is that one cannot have the dominant mass component confined to a narrow range of azimuth in the spiral arms, since the azimuthal velocities are different 
by less than a factor 2 between arm and interarm regions. The lifetime estimate above is of course the mean \underline{molecular lifetime} -- it is not the lifetime 
of the cloud structures, which may well break up into smaller molecular clouds upon leaving the spiral arms. (This issue is discussed in more quantitative detail in \cite{kod16})

\smallskip
\underline{Why is SF concentrated in the spiral arms?} If the H$_2$ clouds have a long lifespan and persist into the interarm regions, why is the visible SF so beautifully concentrated in the spiral arms of M74 and other local galaxies (see Fig. \ref{m74})? Despite the fact that the GMCs are self-gravitating, they are not individually on the verge of collapse to form star clusters due to the supersonic turbulent pressure support, corresponding to a few km s${\rm ^{-1}}$. However, within the spiral arms,
the galactic orbits converge and the cloud-cloud velocity dispersions are increased; cloud-cloud collisions can then compress the internal motions. The molecular gas is very dissipative of the shock energy, leading to collapse of large masses, and in some cases precipitating massive cluster formation. At high redshift gas rich galaxies there may not be such well-organized spiral structure, but there will be a much higher rate of cloud collisions, simply due to the larger number density of cloud structures and the large dispersive motions. A similar scenario probably occurs in the nuclear regions of the local ultraluminous IR galaxies which have much higher SFEs. (The role of GMC collisions in triggering the formation of star clusters is discussed in \cite{sco79,sco87a,fuk14,fuk15,tan00,wub17}.)

\smallskip
\underline{Why is the overall cycling time so long} for galaxies like the Milky Way? Taking the molecular gas content of the 
Milky Way and dividing by the overall SFR of a few \msun per yr yields a very long cycling time of $\sim$ 2 Gyr. Why is this mean SFE so low? Here one should 
recall the internal structure of the GMCs. As mentioned above, the gas is in filamentary structures and when two such clouds collide the filaments in each 
are unlikely to be aligned -- thus only a small fraction of the gas content is likely to be compressed into dissipative collapse. A good example of this is  
provided by the two Orion GMCs associated with Orion A and B (see \cite{lom14}). These elongated clouds have their massive star clusters (M42 and NGC 2024) in the nearest regions of their respective clouds, where they might have collided $\sim$10 Myrs in the past. Very possibly, the low Galactic efficiency for forming stars is due to the internal {\bf filamentary} structure of the molecular clouds. The filamentary structure is also likely to reduce the effectiveness of energetic SF activity in disrupting the clouds -- thus extending their lifetimes. 

In summary of the discussion here, we have pointed out that in galaxies where the SF molecular gas is abundant, 
the molecular gas, and perhaps the clouds are likely to be long-lived, and that a significant fraction of the SF is likely to be associated and triggered by 
compressive collisions of the clouds. Both of these phenomena are likely even more true in the early universe where the gas densities are
much higher and the cloud motions are more disordered.

\section{Comments and Implications}

The variations of gas masses, accretion and their relation to star formation have been explored with the most extensive sample yet of 
high redshift galaxies. Although the deduced estimates are 'consistent' with most existing studies using the CO lines \citep{tac10,dad10,gen10,rie11,ivi11,mag12a,sai13,car13,tac13,bol15,gen15}. However, the sample of galaxies used here is vastly more extensive and has the virtue that it maps the parameter 
space of z, $M_{\rm *}$, and sSFR out to z = 5 using high quality and uniform ancillary data from the COSMOS survey field. We thus can simultaneously constrain the functional dependencies on redshift, sSFR relative to the MS and stellar mass. Our technique also does not suffer from the uncertainties introduced by 
variable excitation in the higher-J CO lines and the dissociation of CO in strong radiation fields. By contrast the dust is much more harder to destroy in strong radiation fields and is a 1\% mass tracer as compared with CO for which the abundance is $\lesssim 10^{-4}$.

A major uncertainty for both the dust and the CO line studies is, of course, their dependence on metallicity (Z). Both are probably more robust than
is generally assumed. The CO line is heavily saturated; in Galactic GMCs which have typical $\rm \tau_{CO (1-0)} \gtrsim 10$. The $^{13}$CO emission 
line is typically $\sim1/5$ of the CO line flux in Galactic GMCs despite the much lower $^{13}$C/C abundance ($\sim$1/60 to 1/90). Thus, the line luminosity must scale 
as the $\sim$1/3 power of the CO abundance, and hence the metallicity \citep[see][for a discussion of excitation by line photon trapping in optically thick lines]{sco74}. 

With respect to the dust emission as a probe of gas, it is reassuring that the dust-to-gas abundance ratio in low redshift galaxies is approximately constant at $\sim$1\% by mass from solar down to 1/5 solar metallicity \citep[see][]{dra07a} and \cite[][Figure 16]{ber16} although why this is the
case is not understood.

Our finding that the gas-to-stellar mass ratio and the accretion rates are both generally higher for lower mass galaxies has implications 
for the gas-phase metallicites of galaxies. Assuming that the metallicity of freshly accreting gas is significantly lower than that of the internal gas in the galaxies, one would expect the gas phase metallicity to increase in higher stellar mass galaxies. This is, of course, known to be true; and it is
a major motivation for our focus on galaxies with relatively high M$_{\rm *}$. It is also clear that a so-called `closed box' model for the evolution of metal content has little 
physical justification in light of the extremely large accretion, SF and feedback rates.

\acknowledgments
Nick Scoville wishes to thank his mentors (Phil Solomon and Peter Goldreich) and his colleagues who have made a career in Astronomy and Astrophysics so  enjoyable. Each step in a new direction has been stimulating, with new insights and unsolved problems over the broad range 
of exploration. The support of the public through funding and their interest and appreciation of the expanding understanding are a constant encouragement.
\medskip

We thank Zara Scoville for proof reading the manuscript. 
ALMA is a partnership of ESO (representing its member states), NSF (USA) and NINS (Japan), together with NRC (Canada), NSC and ASIAA (Taiwan), and KASI (Republic of Korea), in cooperation with the Republic of Chile. 
  The Joint ALMA Observatory is operated by ESO, AUI/NRAO and NAOJ. The National Radio Astronomy Observatory is a facility of the National Science Foundation operated under cooperative agreement by Associated Universities, Inc. ST acknowledges support from the European Research Council (ERC) in the form of Consolidator Grant, 648179, ConTExt. 
The Cosmic Dawn Center is funded by the Danish National Research Foundation under grant No. 140.

\bibliography{scoville_dust.bib}{}

\eject

 \appendix

\section{Long Wavelength Dust Continuum as an ISM Mass Tracer}\label{dust_app}

In this appendix we summarize the physical principles behind our use of  the long wavelength dust continuum 
as a tracer of gas mass. An empirical calibration of this technique is derived from a diverse sample of 128 galaxies at both low redshift and out to z = 4, all having 
global measurements of the long wavelength dust continuum and the well-calibrated CO(1-0) line luminosities (see Fig. \ref{empir_cal}). 

\subsection{A simple physical model for the Dust Radiative Equilibrium and Radiative Transfer}\label{model}

\begin{figure}[ht]
\epsscale{1.2} 
\plotone{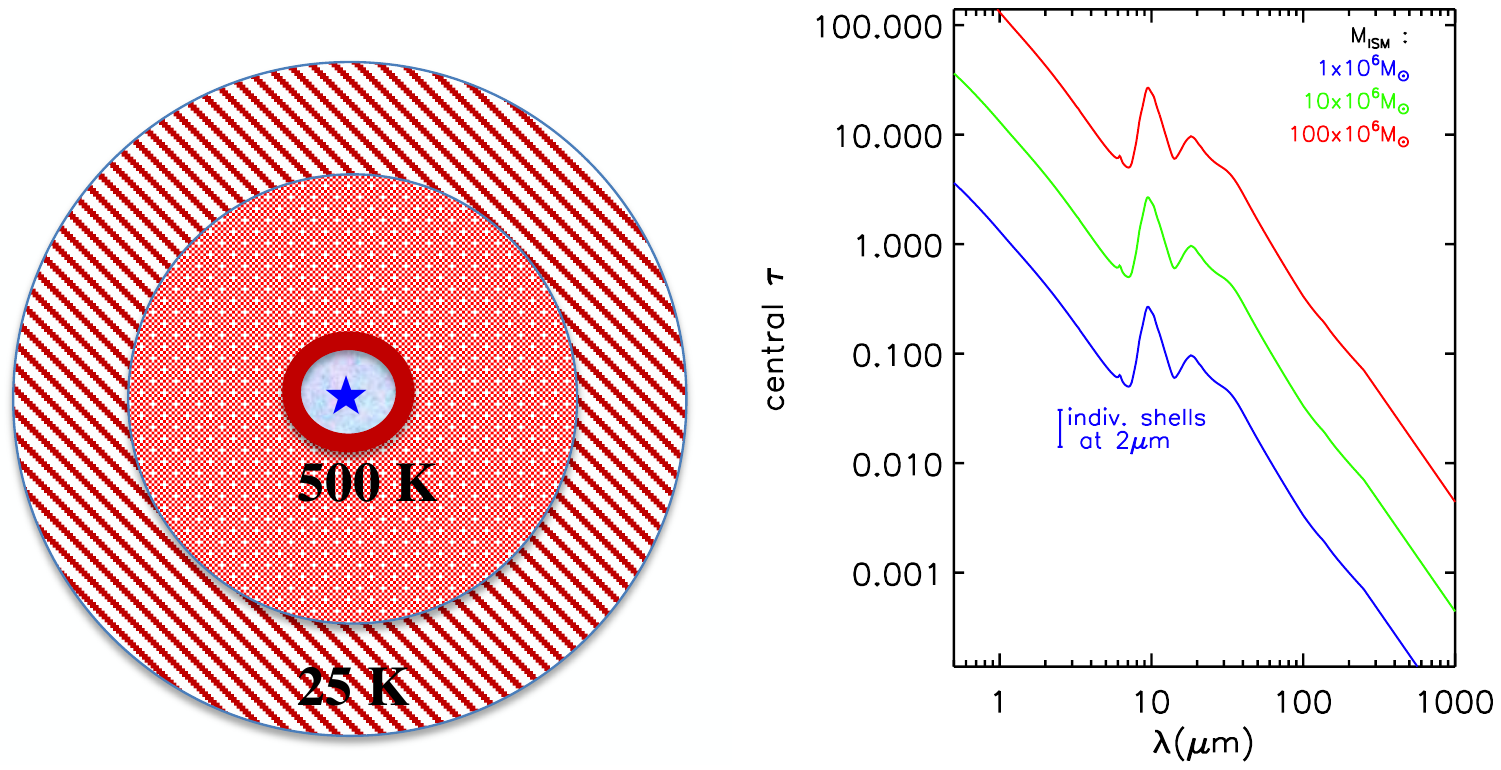} 
\caption{{\bf Left :} Our simple IR radiative heating and transfer model is calculated for a massive molecular gas and dust cloud with a $r^{-1} $ density falloff out to 100 pc radius. At the center of the spherically symmetric cloud, a central luminosity source of $10^7$ \lsun (a young star cluster or AGN, provides the dominant dust heating. Radiative heating from the exterior CMB is also included. The primary source radiation at short wavelengths is absorbed at the innermost radii, and successive shells at larger radii are then heated primarily by secondary and tertiary photons emitted by this lower temperature dust and the CMB. {\bf Right :} The dust optical depths for the three massive molecular gas and dust clouds  are shown integrating in radius from the center to the surface for 
$10^6 - 10^8 $\msun cloud masses, with a $r^{-1} $ density falloff out to 100 pc radius. The dust opacity variation with wavelength is adopted from Draine's absorption coefficient for Milky Way dust. (The blue error bar shows the range of $2 \mu$m optical depths for individual shells in the $10^6$ \msun model.)}
\label{dust_tau} 
\end{figure}

To illustrate the essential physics of the IR emission from dense molecular and dust cloud, we have computed the radiative equilibrium for a spherical dust cloud surrounding a central luminosity source, either a young star cluster or an AGN. In the calculation discussed below we specify 
the dust and gas in 15 shells logarithmically spaced in radius and with density decreasing as $r^{-1}$ out to an outer radius at 100 pc (as shown schematically in Fig. \ref{dust_tau}-left). The central luminosity (presumably visible and UV photons) is taken as a point source blackbody ($T = 3\times10^{4}$K) with $L  = 10^{7} $ \lsun. This luminosity source is surrounded by ionized gas out to $\sim$1 pc, at which point the optical and NUV photons will be absorbed in a thin boundary dust shell, and reradiated in the MID IR. To illustrate the effect of increasing cloud masses (and increasing dust optical depth), the total 
cloud masses were taken to be  $10^6$, $10^7$ and $10^8$ \msun - thus spanning a factor 100 in optical depth
These parameters are chosen to be similar to those 
of massive clouds which might exist in galaxies at z = 2 -3 when there were generally much greater gas masses, 
or the massive gas concentrations in starburst galaxy nuclei. 

For the dust opacity we adopt Bruce Draine's Milky Way dust absorption curve taken from his website file 'Kext-albedo-WD-MW-3.1-60-DO3-all' 
(Fig. \ref{dust_tau}-right) .The scattering and absorption contributions in his table were summed at each wavelength to yield the $\kappa$ values at 517 wavelengths from $\lambda$ = 0.1 to 5000 $\mu$m. (At $\lambda \geq 250$ $\mu$m we use a simple power law with spectral index $\beta = 2$, as expected from the Kramers-Kronig relations at long wavelengths.)  To translate from dust to gas masses, we adopt a constant gas-to-dust mass ratio of 105:1 \citep{dra11}. 

In our numerical integration of the radiative equilibrium and radiation transfer, the central input luminosity is conserved to within $\sim$20\% by the emergent luminosity calculated at r = 100 pc. This is true even in the most optically thick model (e.g. having $10^8$ \msun within r = 100 pc and radial $\tau_{100\mu m} \simeq 1$ as shown in Fig. \ref{cmb_corr}).

\begin{figure} [ht]
\epsscale{0.8} 
\plotone{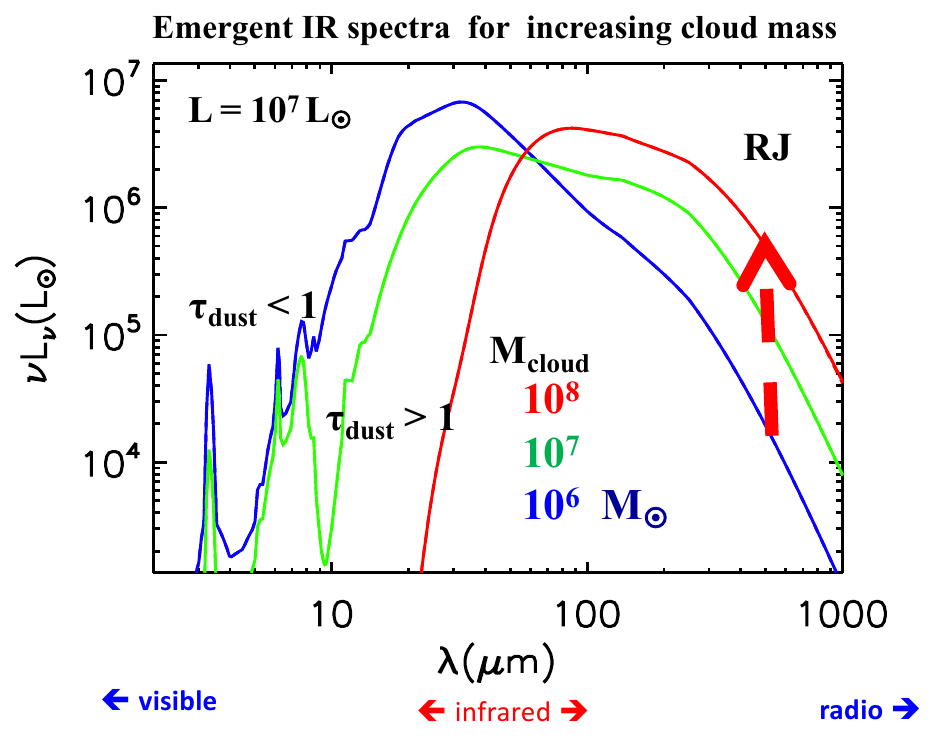} 
\caption{The emergent IR spectra are shown for massive clouds surrounding a central source of $10^7$ \lsun  for masses (gas + dust) of
$10^{~6 ~\rightarrow ~8} $\msun. As the mass increases, the short wavelength IR SED is attenuated and the IR peak shifts to longer wavelengths but the RJ continuum at $\lambda \geq 250 \mu$m increases approximately linearly with increasing overlying mass. At each wavelength, the dust photosphere must be at $\tau_{\lambda} \leq 1$, hence at short wavelengths in high optical depth models one doesn't see sufficiently deeply into the cloud to sample the hot dust at the innermost radii. }
\label{model_spectra} 
\end{figure}

\begin{figure}[ht]
\epsscale{1.0}  
\plotone{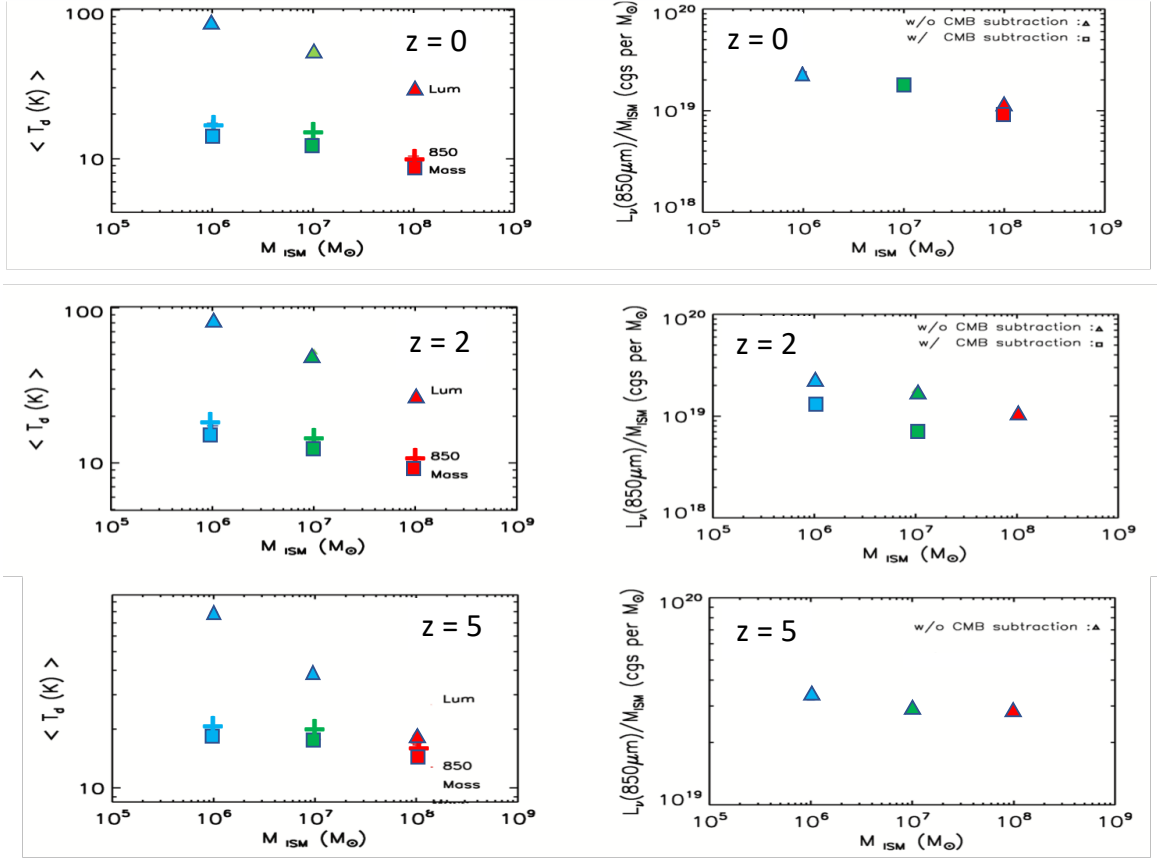}
\caption{{\bf Left Panels :} For z = 0, 2, and 5 the luminosity, mass and 850 $\mu$m flux -weighted dust temperatures for the emergent luminosity are shown for model clouds of $\rm 10^6, 10^7, and~10^8 $\msun from the dust radiative equilibrium model. {\bf Right Panels} show the dependence of the emergent specific luminosity in the source restframe at 850 $\mu$m continuum for the same three masses and redshifts. For the 3 different masses and redshifts, the 850 $mu$m-weighted dust temperature of the emergent luminosity is constant to better than a factor 2 over the full range of a factor 100 in the cloud masses. On the other hand, the luminosity-weighted dust temperature of the emergent radiation varies by a factor $\geq$5. These results are the basis for adopting a constant dust temperature of $\sim 25$ K when using the observed RJ fluxes to estimate the cloud masses out to z = 5 (and not the dust temperature associated with the emission at the FIR peak wavelength). The simple reason for this is that most of the mass in a high optical depth cloud is far away from the central luminosity source -- heated by secondary or tertiary dust photons to only $\simeq 25$ K for the galaxies at $ z \leq 5$ discussed here.}
\label{cmb_corr} 
\end{figure}

\begin{figure*}[ht]
\epsscale{1}  
\plotone{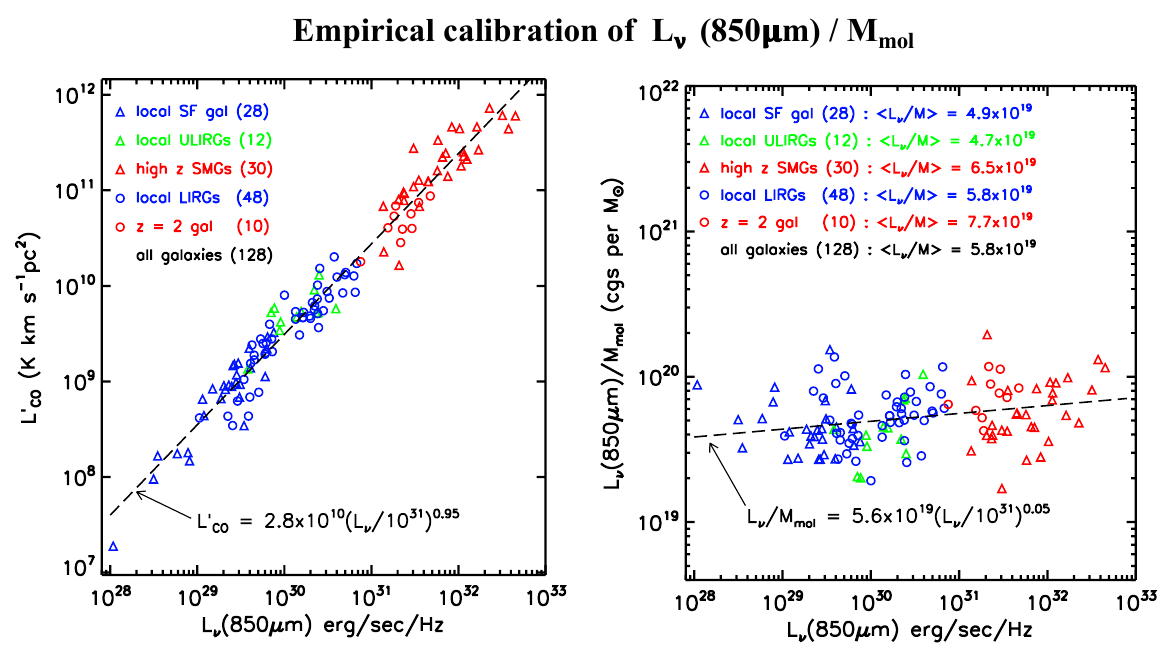}
\caption{{\bf{Left : }} The correlation between the measured CO(1-0) luminosities and the restframe 850 $\mu$m continuum luminosity exhibits a low dispersion over 5 orders of magnitude in luminosity and for a diverse sample of 128 low and high redshift galaxies. .   {\bf{Right: }} The ratio  $L_{\nu}$ at 850$\mu$m to molecular gas masses (derived from CO(1-0) luminosities). 
The galaxy samples are from: 
 low-z star forming) galaxies(28): \cite{You95};
low-z ULIRGs(12): \cite{san89}, \cite{sol97}; 
z = 2 galaxies(10): \cite{kaa19}; 
local LIRGS from Vales Survey(48)): \cite{hug17}, \cite{vil17}; and 
z$\sim$2 SMGs(30): \cite{har12}, \cite{rie11}, \cite{les11}, \cite{tho12} , \cite{tho15}, \cite{ara13}, \cite{ivi11}, \cite{car11}, \cite{gre09}, \cite{tho12}, \cite{har10}, \cite{ivi13} and \cite{fu13}
} 
\label{empir_cal}

\medskip

\end{figure*}

\subsection{Empirical Calibration of the RJ Luminosity to Mass Conversion Factor}\label{empirical}

At long wavelengths on the Rayleigh-Jeans tail, the dust emission is almost always optically thin (see Fig. \ref{dust_tau}-right), and 
the emission flux per unit mass of dust is 
linearly dependent on the dust temperature. Thus, the flux observed on the RJ tail provides a linear estimate (see Fig. \ref{model_spectra}) of the dust mass  and hence the gas mass, provided the dust emissivity per unit mass and the dust-to-gas mass ratio is constrained. 
Fortunately, both of these prerequisites are well established from  observations of nearby galaxies \citep[e.g.][]{dra07b,gal11}. 
 
 On the optically thin RJ tail of the IR emission, the observed flux density is given by 
 
\begin{eqnarray}\label{fnu}
S_{\nu} &\propto&  \kappa_{D}(\nu)  T_{\rm D} \nu^2 {M_{\rm D}\over{d_L^2}} \nonumber 
\end{eqnarray}

\noindent where $T_{\rm D}$ is the temperature of the emitting dust grains,  $\kappa_{D}(\nu)$ is the dust opacity per unit mass of dust, $M_{\rm D}$ is the total
mass of dust and $d_L$ is the source luminosity distance. Thus, the mass of dust and gas can be estimated from observed specific luminosity $L_{\nu}$ on the RJ tail:

\begin{eqnarray}\label{lnu}
M_{dust} &\propto& { L_{\nu}  \over { <T_D>_M   \kappa_{D}(\nu)}}  \\
M_{gas} &\propto& { L_{\nu}  \over { f_{D} <T_D>_M   \kappa_{D}(\nu)}}  .
\end{eqnarray}

\noindent Here $<T_D>_M$ is the mean mass-weighted dust temperature and $f_{D}$ is the dust-to-gas mass ratio (typically $\sim 1/100$ for solar metallicity gas).

 The empirical calibration of the technique is based on 5 different low and high redshift galaxy samples: 1) a sample of 30 local star forming galaxies; 2) 12 low-z Ultraluminous Infrared Galaxies (ULIRGs);  and 3) 30 z $\sim$ 2 submm galaxies (SMGs). These three samples with 128 galaxies are restricted to only those galaxies having good estimates of the 
total source-integrated, long-wavelength continuum and CO (1-0)). (NB: Once these samples were selected for the calibration, no individual galaxies were selectively dropped due to departures from the mean.)

We avoid using higher-J CO lines since only the 1-0 transition has been well-calibrated using large samples of 
Galactic GMCs with viral mass estimates. The emissions in the J = 3-2 and higher CO lines originate only from high excitation (both high density and  heated) cloud core regions. These higher CO lines will not be reliable as overall mass tracers of molecular gas contents. Unfortunately ALMA does not presently have frequency coverage to enable observations of the J =1-0 or 2-1 lines at z $\geq$ 2. There is also no physical justification for observers to claim a fixed ratio of the higher lines to J = 1 - 0 and to thereby infer overall gas contents from the higher J CO emission. The 3-2, 4-3 and 5-4 lines originate from levels at $E_u/k$ = 33, 55 and 82 K above the ground state, whereas most of the cloud mass is expected to be at $\sim 25$ K. The higher J lines originate largely from compact core regions, in contrast to the 1-0 line, which arises from the full cloud extent, and thus samples the overall mass.

In calibrating the CO(1-0) masses, we have adopted $\alpha_{CO(1-0)}  = 3\times10^{20} cm^{-2} (K \kms)^{-1}$ which 
is derived from correlation of the CO line luminosities and virial masses for resolved Galactic GMCs. We believe this is more correct than the value obtained from Galactic gamma ray 
surveys ($\alpha = 2\times10^{20}$) \citep[see][]{bol13}, since the latter requires questionable assumptions: 1) that the cosmic rays which produce the $\sim2$ MeV gamma rays by interaction with the gas fully penetrate the GMCs and 2) 
the cosmic ray density with Galactic radius. (If one adopted the latter value of $\alpha_{CO}$, the derived scaling for the dust-based gas masses would be reduced by a factor of 2/3.)

All galaxies in our RJ  calibration sample were required to have {\bf global measurements of CO (1-0) and Rayleigh-Jeans dust continuum}. The large range in apparent luminosities at 850$\mu$m and in CO is due to the inclusion  of high redshift SMGs, many of which 
are strongly lensed. The samples are all processed in a common way : 

1) All molecular gas masses were derived using the same CO (1-0) conversion factor $X_{CO} = 3\times10^{20}$ N(H$_2$) cm$^{-2}$ (K km s$^{-1})^{-1}$; 
2) The molecular gas masses all include a correction for the associated masses of He.
 
 These diverse samples yield remarkably similar values for the dust-to-gas conversion factor $\rm \alpha_{850} = 5.6\times10^{19}$ cgs per \msun (see dashed line fit in Fig.  \ref{empir_cal} - Right).
The Planck value for the Milky Way converted to the same  $X_{CO}$ is 6.2$\rm \times10^{19}$ cgs per \msun 
between the dust continuum flux and the molecular masses (including He).
We have adopted a long wavelength spectral index for the dust opacity of $\beta_{dust}=2$ for all these datasets
when converting the RJ flux at one wavelength to the reference $\lambda = 850 \mu m$.

Figure \ref{empir_cal} shows the ratio of specific luminosity at rest frame $\lambda = 850 \mu$m to that of the CO (1-0) line, and one clearly sees a similar ratio
of RJ dust continuum to CO luminosity. Using a standard Galactic CO (1-0) conversion factor, we then obtain the relation by which we convert the RJ dust continuum to gas masses:

  \begin{eqnarray}
M_{\rm gas}   &=& 1.78 ~S_{\nu_{obs}}[\rm mJy]  ~ (1+z)^{-4.8} ~   \left({ \nu_{850\mu \rm m}\over{\nu_{obs} }}\right)^{4.0}   ({\it{d}_L \rm{[Gpc ]}})^{2} \nonumber \\
 && \times ~\left\{{5.6\times10^{19}\over{\alpha_{850}  }} \right\}    ~{\it{\Gamma_{0}} \over{\it{\Gamma_{RJ}}}}~   ~10^{10}\msun    ~. \label{mass_eq}  \label{dust_eq}
  \end{eqnarray}
  
\begin{figure}[ht]
\epsscale{1.2}  
\plotone{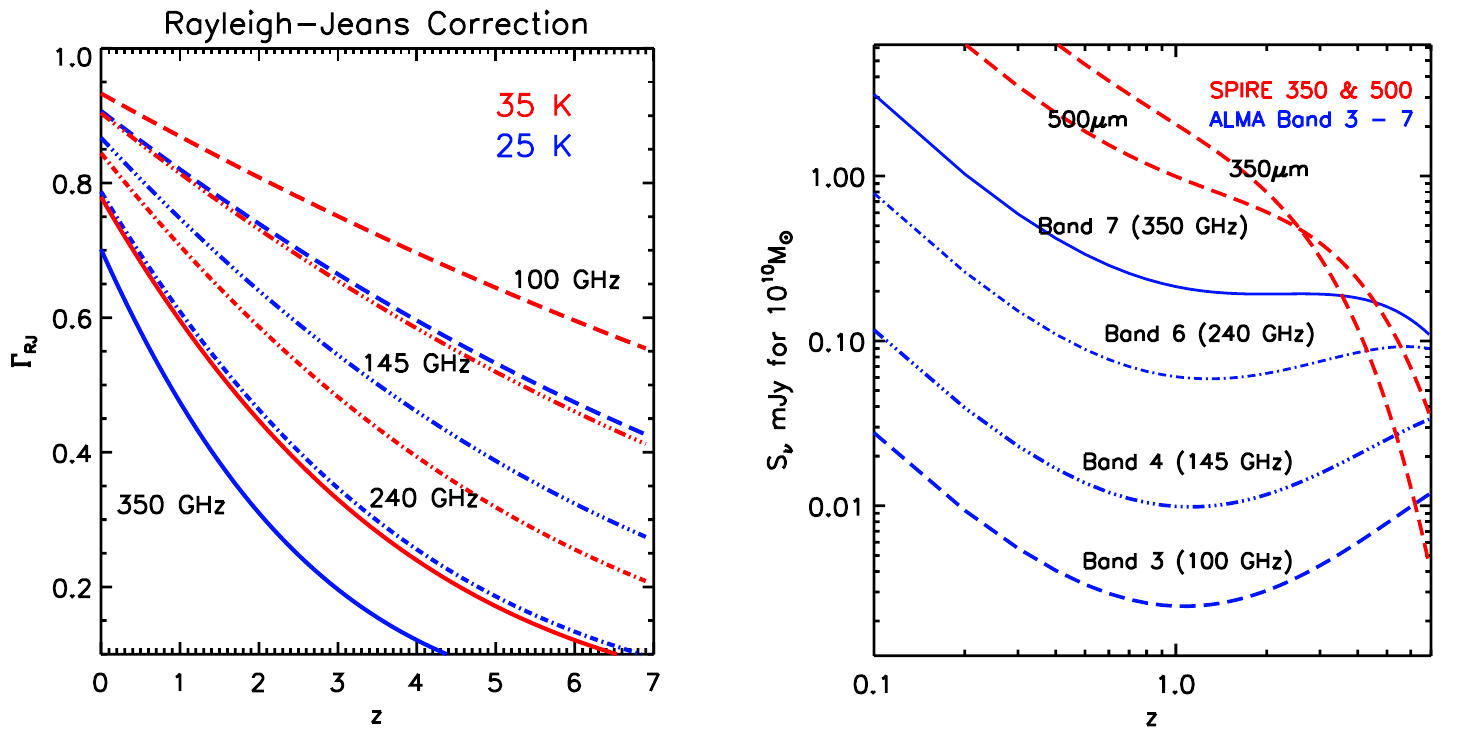}
\caption{{\bf Left:} The correction to the Rayleigh-Jeans continuum \citep{sco17} is plotted as a function of z for observed frequencies 100 - 350 GHz. (The correction factor of course becomes very large as one observes at high z and is no longer on RJ side of the emission. One should be observing a lower frequency band.) {\bf Right:}  The expected continuum flux densities expected for the ALMA bands at 100, 145, 240 and 350 GHz and for SPIRE 350 and 500$\mu$m for a fiducial M$_{gas} = 10^{10}$\msun ~derived using the empirical calibration $\alpha = 5.6\times10^{19}$, 
an emissivity power law index $\beta = 2$ and including the RJ departure coefficient $\Gamma_{RJ} (25K)$. Since the point source flux sensitivities of ALMA 
in the 4 bands are quite similar, the optimal strategy is to use Band 7 out to z $\sim 2 -3$; above z = 3. Lower frequency ALMA bands are required to avoid large uncertainties in the RJ correction as shown on the right side of the right figure.}
\label{alma_obs} 
\end{figure}

In Equation  \ref{dust_eq}, $\Gamma$ is a correction for departures from strict $\nu^2$ of the RJ continuum \citep{sco16}, shown in Left panel of Fig. \ref{alma_obs}. $\alpha_{850} = 5.6\pm1.7\times 10^{19} \rm erg ~sec^{-1} Hz^{-1} {\msun}^{-1}$ is the derived calibration constant 
between 850$\mu$m luminosity and gas mass. We have adopted a dust opacity spectral index $\beta = 2$ (see \S \ref{model}). 

 In the present work, the conversion of the fluxes, measured with ALMA, to gas masses is done with Equation \ref{dust_eq}. Using Equation \ref{dust_eq}, the predicted fluxes for a fiducial gas mass of $10^{10}$\msun ~in the ALMA and Herschel-SPIRE bands are shown in Figure \ref{alma_obs} as a function 
of redshift.

In the current work we restrict the observed galaxies to be relatively massive ($M_{*} > 5\times10^{10}$ \msun) since they should have close to solar metallicity and presumably close to solar dust-to-gas abundance ratios. We note that for the first factor $\sim5$ down from solar metallicity the dust-to-gas ratio is constant for those galaxies with global measurements \citep{dra07b}.   Probing lower stellar mass galaxies, which presumably would have significantly sub-solar metallicity, will require careful calibration as a function of metallicity or mass. We note that in \cite{dra07b}, there is little evidence of variation in the dust-to-gas abundance ratio for the first factor of 4-5 down from solar metallicity. However, at lower metallicities the 
dust-to-gas abundance does clearly decrease (see Figure 17 in \cite{dra07b} and Figure 16 in \cite{ber16}). 

\subsection{A Caution for higher redshifts}

The effects of increased CMB \citep{da_Cunha13} at higher redshifts were included, but these effects are only significant at z $\geq 5$. There were 
two effects pointed out by \cite{da_Cunha13} : increased heating due to the higher CMB temperature, and decreased CMB flux in the 
source position (compared to the sky reference position which inevitably is subtracted in the differential observations), due to dust absorption of the CMB passing through the galaxy. The first effect is easily accounted for in our model since the 
CMB flux penetrating to each location in the cloud is taken into account in the dust heating. Unfortunately, the second effect, where too much CMB 
is assumed in the reference beam, compared to the on-source beam, requires knowing the spatial distribution of dust opacity within the observed sources, or assuming the dust is in the low opacity (linear) limit, which is probably not always true for z $\geq 5 $ massive star forming galaxies. It is possible that this second effect might be corrected partially by dual frequency flux measures, but that is beyond the scope of the efforts here and is not needed for the redshift range considered here.

\section{Continuity of the MS Evolution} \label{continuity}

In our analysis, we make use of a principle we refer to as the \emph{Continuity of Main Sequence Evolution} -- simply stated,  
the temporal evolution of the SF galaxy population may be followed by Lagrangian integration in the $SFR/M_*$ plane. This follows from the fact that
approximately 95\% of SF galaxies at each epoch lie on the MS with SFRs dispersed only a factor 2 above or below the MS \citep{rod11}. A similar approach has been used by 
\cite{noe07a}, \cite{ren09}, and \cite{{lei12}}, and references therein.

This continuity assumption ignores the galaxy buildup arising from \emph{major} mergers of similar mass galaxies since they can depopulate the MS population in the mass range of interest.  
Major mergers may be responsible for some galaxies in the SB population above the MS (see Section \ref{sb_ms}). On the other hand, minor mergers may be considered simply as one element of the average accretion process considered in Section \ref{accretion}. 

We are also neglecting the SF quenching processes in galaxies. This occurs mainly in the highest mass galaxies (M$_{\rm *} > 2\times10^{11}$\msun ~ at z  $>1$) and in dense environments at lower redshift \citep{pen10,dar16}. At z $> 1.2$ the quenched red galaxies 
are a minor population (see Figure \ref{lilly_madau}) and the quenching processes are of lesser importance.

The paths of evolution in the 
SFR versus M$_{\rm *}$ plane can be easily derived  since the MS loci give $dM_{\rm *}/dt$ = SFR (M$_{\rm *}$). 
One simply follows each galaxy in a Lagrangian fashion 
as it builds up its mass. In the Lagrangian integration, we move with the galaxies as they trace these paths, and the time derivatives of a mass component M are taken along the evolutionary track. Fig. \ref{evolution_tracks} shows these evolutionary tracks for a sample of galaxies.
Using this Continuity Principle to evolve each individual galaxy over time, the evolution for MS galaxies across the SFR-M$_{\rm *}$ plane is as shown in Figure \ref{evolution_tracks}. Here we have assumed that 30\% of the SFR is eventually put back into the gas by stellar mass-loss. This is appropriate for the mass-loss from a stellar population with a Chabrier IMF \citep[see][]{lei11}.
In this figure, the curved horizontal lines are the MS at fiducial epochs or redshifts, while the downward curves are the evolutionary tracks 
for fiducial M$_{\rm *}$ from 1 to 10 $\times10^{10}$ \msun. At higher redshift, the evolution tends toward increasing M$_{\rm *}$ whereas at lower redshift the evolution is  
in both SFR and M$_{\rm *}$. In future epochs, the evolution is more vertical as the galaxies exhaust their gas supplies. Thus there are three phases in the evolution:
\begin{enumerate}
\item the gas accretion-dominated and stellar mass buildup phase at z $ > 2 $ corresponding to cosmic age less than 3.3 Gyr (see Section \ref{accretion});
\item the transition phase where gas accretion approximately balances SF consumption and the evolution becomes diagonal;
and
\item the epoch of gas exhaustion at z $\lesssim$ 0.1 (age 12.5 Gyr) where the evolution is vertically downward in the SFR versus M$_{\rm *}$ plane.
\end{enumerate} 

These evolutionary phases are all obvious (and not a new development here). However, in Section \ref{accretion} we make use of this Continuity assumption to derive the accretion rates and hence substantiate the 3 phases as separated by their accretion rates relative to their SFRs. When 
these phases begin and end is a function of the galaxy stellar mass -- the transitions in the relative accretion rates take place at higher redshift (i.e. earlier cosmic epoch) for the more massive galaxies. Figure \ref{evolution_tracks} shows the evolution for three fiducial stellar masses. 
 

At each epoch there exists a much smaller population 
($\sim5$\% by number at z = 2) which has SFRs 2 to 100 times that of the MS at the same stellar mass. Do these starburst galaxies quickly exhaust their supply of star forming gas, thus evolving rapidly back to the MS? Or are their gas masses systematically larger so that their depletion times differ little from the MS galaxies? These SB galaxies must be either a short-duration, but common, evolutionary phase or of long-duration but not undergone by the majority of the galaxy population. Their significance in the overall 
cosmic evolution of SF is certainly greater than 5\%, since they have 2 to 50 times higher SFRs than the MS galaxies.

\section{Datasets and Measurements}\label{datasets}

The major datasets used for our analysis were : 
\begin{enumerate} 

\item the latest COSMOS 2020 photometric redshift catalog \citep{wea22} for positions, stellar masses and the unobscured SFRs; 

\item  the far infrared continuum fluxes from Herschel PACS and SPIRE for the IR-based obscured SFR rates; and

\item ALMA bands 6 and 7 continuum imaging for estimating gas masses from the long wavelength RJ fluxes.

\end{enumerate}

\subsection{Redshifts, Stellar Masses and SFRs : Optical/UV and IR Star Formation Rates}
 
Photometric redshifts from \cite{wea22} were used for all sources. Our final catalog does 
not include objects for which the photometric redshift fitting, Xray or radio emission  indicated a possible AGN. 
The primary motivation for using the Herschel IR catalogs for positional priors is the fact that once one has far-infrared detections of a galaxy, the SFRs can be estimated more reliably (including the dominant 
contributions of dust-obscured SF) rather than relying on optical/UV continuum estimations, which often have  extinction corrections by factors $\gg5$ for dust obscuration.  This said, the SFRs derived from the far-infrared are still individually uncertain by a factor 2, given uncertainties in the stellar IMF and the assumed timescale over which young stars remain dust-embedded. 

 The conversion from IR (8-1000$\mu$m) luminosity makes use of $\rm SFR_{IR} = 8.6\times10^{-11} L_{IR}/\lsun$ using a Chabrier stellar IMF from 0.1 to 100 \msun \citep{cha03}.
The scale constant is equivalent to assuming that 100\% of the stellar luminosity is absorbed by dust for the 
first $\sim$100 Myr and 0\% for later ages. For a shorter dust enshrouded timescale of 10 Myr the scaling constant is $\sim$1.5 times larger \citep{sco13a},
so this duration of dust obscuration is not a critical uncertainty. In 706 of the 708 sources in our measured sample, the IR SFR was greater than the optical/UV SFR. The final SFRs are the sum of the opt/UV (with extinction corrections removed) and the IR SFRs.  

The stellar masses of the galaxies are taken from the latest COSMOS 2020 photometric redshift catalog \citep{wea22} and the lower limit for the stellar masses was $\sim 5\times 10^9$ \msun; the $\rm M_*$  are  
probably  uncertain in some instances by a  factor of 2 due to uncertainties in the spectral energy distribution (SED) modeling and extinction corrections. Their uncertainties are less than those for the optically derived SFRs, since the stellar mass in galaxies is 
typically more extended than the SF activity, and therefore is likely to be less extincted. 

The other galaxy property we wish to correlate with the derived gas masses is 
the elevation of the galaxy above or below the SF MS. This enhanced SFR is quantified by sSFR/sSFR$_{\rm MS}(z,M_{\rm *})$ with the MS definition taken from the combination of \cite{spe14} and \cite{lee15}.  

\subsection{IR Source Catalogs}\label{ir_data}

Our source finding used a positional prior: the Herschel-based catalog of far-infrared sources in the COSMOS field \citep[13597 sources]{lee13,lee15}. COSMOS was observed at 100 $\mu$m and 160 $\mu$m by Herschel PACS \citep{pog10} as part of the PACS Evolutionary Probe program \citep[PEP;][]{lut11}), and down to the confusion limit at 250 $\mu$m, 350 $\mu$m, and 500 $\mu$m by Herschel SPIRE \citep{gri10} as part of the Herschel Multi-tiered Extragalactic Survey  \citep[HerMES;][]{oli12}). 

In order to measure accurate flux densities of sources in the confusion-dominated SPIRE mosaics, it is necessary to extract fluxes using prior-based methods, as described in \cite{lee13}. In short, we begin with a prior catalog that contains all COSMOS sources detected in the Spitzer 24$\mu$m and VLA 1.4 GHz catalogs \citep{lef09,sch10}, which have excellent astrometry. Herschel fluxes are then measured using these positions as priors. The PACS 100 and 160 $\mu$m prior-based fluxes were provided as part of the PEP survey \citep{lut11}, while the SPIRE  250, 350, and 500 $\mu$m fluxes are measured using the XID code of \cite{ros10,ros12}, which uses a linear inversion technique of cross-identification to fit the flux density of all known sources simultaneously \citep{lee13}. From this overall catalog of infrared sources, we select reliable far-infrared bright sources by requiring at least 3$\sigma$ detections in at least 2 of the 5 Herschel bands. This greatly limits  the number of false positive sources in the catalog.

An in-depth analysis of the selection function for this particular catalog is provided in \cite{lee13}, but the primary selection function is set by the 24 $\mu$m and VLA priors catalog. As with many infrared-based catalogs, there is a bias toward bright, star-forming galaxies, but the requirement of detections in multiple far-infrared bands leads to a flatter dust temperature selection function than typically seen in single-band selections. 

For the IR luminosities we use only sources listed in at least 2 of the 3 IR catalogs for COSMOS. One of these was the Sextractor catalog of \cite{lee13}.  The other two catalogs use deblending techniques on the Herschel images to go to deeper flux levels and deblend nearby galaxies \citep{hur17,jin18}. Since the latter catalogs may not be entirely reliable we require that they agree or that they are for a source also listed in the Sextractor catalog. All three catalogs used position priors from the Spitzer 24 \mum data of \citep{san07}.

 The IR based SFRs are estimated from the average IR catalog fluxes in each band in two steps : 1) SED fitting and integration over wavelengths, using the code described in 
 \cite{cas12}, and 2) a simple sum of $\nu S_{\nu}$ for all detected bands. The latter is used to eliminate any objects with incomplete IR spectral coverage for the SED fitting. We required that the derived luminosity from the the two techniques must agree to a factor 2 - 3, otherwise the candidate source is dropped.  

Since the selection function is biased to IR bright and massive galaxies, the sample is not representative of lower mass galaxies ($\rm \lesssim 10^9 $ \msun) in the high redshift SF galaxy population. However, in the analytic fitting 
below we obtain analytic dependencies for the gas masses and SFRs on the sSFR, the stellar mass and redshift. These analytic fits are then used to analyze 
the more representative populations. This approach is used in Section \ref{cosmic} to estimate the cosmic evolution of gas, the dependence of gas masses on stellar masses of each galaxy, and the dependence on whether the galaxy was on or above the 
MS at each redshift. 

\subsection{ALMA Bands 6 and 7 Continuum Data}

For the gas mass estimates, we use exclusively continuum observations from ALMA -- these are consistently 
calibrated and with resolution ($\sim1$\arcsec) such that source confusion is not an issue. Lastly, our analysis involving the RJ dust continuum avoids the issue of variable excitation, 
which causes uncertainty when using different CO transitions across galaxy samples. The excitation and brightness per unit mass for the different CO transitions is likely to vary by factors of 2 - 3 from one galaxy to another 
and within individual galaxies \citep[see][]{car13}. 

Within the COSMOS survey field, there now exist extensive observations from ALMA for the dust continuum of high redshift 
galaxies. Here, we make use of all those data which are publicly available as of 6/2021. The number of ALMA pointings in these datasets in Bands 3-7 and 6-7  is 2217, including only ALMA data with uv coverage such that the flux recovery is good out to source extents of $\sim$ 1\arcsec. The Band 6 and 7 
observations (2600 images) used here cover a total area 0.0529 deg$^2$ or 190 arcmin$^2$ within the Half-Power Beam Widths (HPBWs). There were 749 measured fluxes included in ALMA Band 6 and 7 and 708 unique sources. Although the Band 3 - 5 data had some detections, they were not used here since their resulting RJ sensitivity to gas mass was less. The COSMOS survey has a full area of 1.8 deg$^2$ so the actual coverage of COSMOS by ALMA is only 3\%.

In all cases, the ALMA source measures include both a flux measurement and uncertainty estimate for the least squares fitting.
At each IR source position falling within 
the ALMA primary beam HPBW (typically 20\arcsec~in Band 7), we searched for a significant emission source ($> 2 \sigma$) within 2\arcsec~radius of the IR source position. This radius is the expected maximum size for these galaxies. The adopted detection 
limit implies that $\sim2$\% of the detections at the $2\sigma$ limit will be spurious. Since there are $\sim240$ galaxies detected at 2-3$\sigma$, we can expect $\sim4$ of the detections will be false. 

Some of the sources are likely to be somewhat extended relative to the ALMA synthesized beams (typically $\sim1$\arcsec); we therefore measure both the peak and integrated fluxes. The latter were corrected for the fraction of the synthesized beam falling outside the aperture. 
The adopted final flux for each source was the maximum of these as long as the SNR was $>2$. 

The noise for both 
the integrated and peak flux measures is estimated by placing 50 randomly positioned apertures of similar size in other areas of the FOV and measuring the 
dispersion of those measurements. The synthesized beams for most of these observations were $\sim 0.5 - 1$\arcsec, and the interferometry will have good flux 
recovery out to sizes $\sim4$ times this. Since the galaxy sizes are typically $\leq2$ to 3\arcsec, we expect the ALMA flux recovery to be nearly complete; that is, there should be relatively little resolved-out emission.  
All measured fluxes are corrected for primary beam attenuation. The maximum correction is a factor 2 when the source is near the HPBW radius for the 12m telescopes.

A total of 708 of the Herschel sources are found within the ALMA FOVs. The positions of this sample are used as priors for the ALMA flux measurements. For the sample of 708 objects, the measurements yields 708 and 182 objects with $>2$ and $>7\sigma$ detections, respectively. Thus at 2$\sigma$, all sources are detected. No correction for Malmquist bias was applied since 
there were detections for the complete sample of sources falling within the survey area. 
Some of the Herschel sources had multiple 
ALMA observations. 
In summary, all of the Herschel sources within the ALMA pointings were detected. The final sample of 708 galaxies with their redshifts and stellar masses is shown in Figure \ref{sample_m_sfr}. 

 In approximately 10\% of the ALMA images there is more than one detection. However, the redshifts of these secondary sources, and 
the distributions of their offsets from the primary source, indicate that most of the secondaries are not physically associated with the primary IR sources.

\vfill
\eject

\end{document}